\documentclass[useAMS,usenatbib,usegraphicx,onecolumn]{mn2e}

\usepackage{times}
\usepackage{amsmath}
\usepackage{amsfonts}

\title[Bivariate distribution function: application to the LF]{
  Constructing a bivariate distribution function with given marginals and correlation: 
  application to the galaxy luminosity function}

\author[T.\ T.\ Takeuchi]
  {Tsutomu T.~Takeuchi\thanks{E-mail: takeuchi@iar.nagoya-u.ac.jp}\\
  Institute for Advanced Research, Nagoya University, 
  Furo-cho, Chikusa-ku, Nagoya 464--8601, Japan
}

\date{Released 2002 Xxxxx XX}

\newtheorem{theorem}{Theorem}
\newtheorem{definition}{Definition}

\def\pd{{\rm d}}
\def\vmatrix{{\mathbf{\Sigma}}}
\def\imatrix{{\mathbf{I}}}
\def\vx{\mathbf{x}}
\def\vpsi{\mathbf{\Psi^{-1}}}
\def\rs{\rho_{\rm S}}
\def\nc{n_{\rm c}}
\def\nd{n_{\rm d}}
\def\lir{L_{\rm FIR}}
\def\ltir{L_{\rm TIR}}
\def\luv{L_{\rm FUV}}
\def\llim{L^{\rm lim}}
\def\sdet{\Sigma^{\rm det}}
\def\sulo{\Sigma^{\rm UL1}}
\def\sult{\Sigma^{\rm UL2}}
\def\vel{\frac{\displaystyle {\rm d}^2 V}{{\rm d} z' {\rm d} \Omega}}
\def\sfruv{\mbox{SFR}_{\rm FUV}}
\def\sfrir{\mbox{SFR}_{\rm dust}}
\def\sfr{\mbox{SFR}_{\rm tot}}

\begin{document}

\label{firstpage}

\maketitle

\begin{abstract}
We show an analytic method to construct a bivariate distribution function
(DF) with given marginal distributions and correlation coefficient.
We introduce a convenient mathematical tool, called a copula, to connect two 
DFs with any prescribed dependence structure.
If the correlation of two variables is weak (Pearson's correlation coefficient 
$|\rho| <1/3 $), the Farlie-Gumbel-Morgenstern (FGM) copula
provides an intuitive and natural way for constructing such a bivariate DF.
When the linear correlation is stronger, the FGM copula cannot work anymore.
In this case, we propose to use a Gaussian copula, which connects two given
marginals and directly related to the linear correlation coefficient between
two variables.
Using the copulas, we constructed the BLFs and discuss its statistical properties.
Especially, we focused on the FUV--FIR BLF, since these two luminosities are 
related to the star formation (SF) activity.
Though both the FUV and FIR are related to the SF activity, the univariate LFs 
have a very different functional form: former is well described by the Schechter function 
whilst the latter has a much more extended power-law like luminous end.
We constructed the FUV-FIR BLFs by the FGM and Gaussian copulas with different
strength of correlation, and examined their statistical properties.
Then, we discuss some further possible applications of the BLF: the problem of 
a multiband flux-limited sample selection, the construction of the SF rate (SFR)
function, and the construction of the stellar mass of galaxies ($M_*$)--specific SFR
($\mbox{SFR}/M_*$) relation.
The copulas turned out to be a very useful tool to investigate all these issues, 
especially for including the complicated selection effects.
\end{abstract}


\section{Introduction}

A luminosity function (LF) of galaxies is one of the fundamental tools to describe 
and explore the distribution of luminous matter in the Universe.
\citep[see, e.g.][]{binggeli88,lin96,takeuchi00a,takeuchi00b,blanton01,delapparent03,willmer06}.
Up to now, studies on the LFs have been rather restricted to a univariate one, i.e.\ LFs
based on a single selection wavelength band.
However, such a situation is drastically changing in the era of large and/or deep Legacy surveys.
Indeed, a vast number of recent studies are multiband-oriented: they require data from 
various wavelengths from the ultraviolet (UV) to the infrared (IR) and radio bands.
A bivariate LF (BLF) would be a very convenient tool in such studies.
However, to date, it is often defined and used in a confused manner, 
without careful consideration of complicated selection effects in both bands.
This confusion might be partially because of the intrinsically complicated nature of
multiband surveys, but also because of the lack of proper recipes to describe a BLF.
Then, the situation will be remedied if we have a proper analytic BLF model.

However, it is not a trivial task to determine the corresponding bivariate 
function from its marginal distributions, if the distribution is not multivariate 
Gaussian.
In fact, there exist infinitely many distributions with the same marginals because the 
correlation structure is not specified.
In general astronomical applications (not only BLFs), for instance, 
a bivariate distribution is often obtained by either an {\it ad hoc} or a heuristic manner
\citep[e.g.][]{choloniewski85,chapman03,schafer07}, though these 
methods are quite well designed in their purposes.
Further, analytic bivariate distribution models are often required to
interpret the distributions obtained by nonparametric methods
\citep[e.g.][]{cross02,ball06,driver06}].
For such purposes, a general method to construct a bivariate 
distribution function with pre-defined marginal distributions 
and correlation coefficient is desired.

In econometrics and mathematical finance, such a function has been commonly 
used to analyze two covariate random variables.
This is called ``copula''.
Especially in a bivariate context, copulas are useful to define nonparametric 
measures of dependence for pairs of random variables \citep[e.g.][]{trivedi05}.
In astrophysics, however, it is only recently that copulas attract researchers'
attention and are not very widely known yet 
\citep[still only a handful of astrophysical applications:][]{benabed09,jiang09,koen09,scherrer10}.
Hence the usefulness and limitations of copulas are still not well understood in the
astrophysical community.

In this paper, we first introduce a relatively rigorous definition of a copula.
Then, we choose two specific copulas, the Farlie-Gumbel-Morgenstern (FGM) 
copula and the Gaussian copula to adopt for the construction of a model BLF.
Both of them have an ideal property that they are explicitly related to
the linear correlation coefficient.
Though, as we show in the following, the linear correlation coefficient is not a
perfect measure of the dependence of two quantities, this is the most familiar
and thus fundamental statistical tool for physical scientists.
We focus on the far-infrared (FIR)--far-ultraviolet (FUV) BLF as a concrete 
example, and discuss its properties and some applications.

This paper is organized as follows: in Section~\ref{sec:formulation} we define a copula and
present its dependence measures. We also introduce two concrete functional forms, 
the Farlie-Gumbel-Morgenstern copula and the Gaussian copulas.
In Section~\ref{sec:blf}, we make use of these copulas to construct a BLF of
galaxies. Especially we emphasize the FIR-FUV BLF.
We discuss some implications and further applications in Section~\ref{sec:discussion}.
Section~\ref{sec:conclusion} is devoted to summary and conclusions.
In Appendix~\ref{sec:jk77}, we show an iterated extension of the FGM copula.
We present statistical estimators of the dependence measures of two variables 
in Appendix~\ref{sec:nonparam_est} to complete the discussion.

Throughout this paper, we adopt a cosmological model $(h, \Omega_{\rm M0}, \Omega_{\rm \Lambda0})
= (0.7, 0.3, 0.7)$ ($h \equiv H_0/100 [\mbox{km}\,\mbox{s}^{-1}]$) unless otherwise stated.

\section{Formulation}
\label{sec:formulation}

\subsection{Copula}
\label{subsec:copula}

As we discussed in Introduction, there is an infinite degree of freedom to choose
a dependence structure of two variables with given marginal distribution.
However, very often we need a systematic procedure to construct a bivariate
distribution function (DF) of two variables\footnote{In this work, we use a term DF 
for a cumulative distribution function (CDF). 
We use a term probability density function (PDF) to avoid confusion with the term
used in physics ``distribution function'' which stands for a Radon-Nikodym
derivative of a DF.}.
Copulas have a very desirable property from this point of view. 
In short, copulas are functions that join multivariate DFs to their one-dimensional
marginal DFs.
However, this statement does not serve as a definition. 
We first introduce its abstract framework, and move on to a more concrete form which
is suitable for the aim of this work (and of many other physical studies).

Before defining the copula, we prepare some mathematical concepts in the following.
\begin{definition}
Let $S_1$ and $S_2$ be nonempty subsets of $\bar{\mathbb{R}}$
[a union of real number $\mathbb{R}$ and $(-\infty, \infty)$].
Let $H$ be a real function with two arguments (referred to as bivariate or 2-place) 
such that $\mbox{Dom } H = S_1 \times S_2$.
Let $B = [x_1, x_2] \times [y_1, y_2]$ be a rectangle all of whose vertices are in
$\mbox{Dom } H$.
Then, the {\it $H$-volume} of $B$ is defined by 
\begin{eqnarray}
  V_H (B) \equiv H(x_2,y_2) - H(x_2,y_1) - H(x_1, y_2) + H(x_1,y_1) \;.
\end{eqnarray}
\end{definition}
\begin{definition}
A bivariate real function $H$ is {\it 2-increasing} if $V_H \geq 0$
for all rectangles $B$ whose vertices lie in $\mbox{Dom } H$.
\end{definition}
\begin{definition}
  Suppose $S_1$ has a least element $a_1$ and $S_2$ has a least element $a_2$.
  then, a function $H$: $S_1 \times S_2 \rightarrow \mathbb{R}$ is {\it grounded}
  if $H(x,a_2) = 0$ and $H(a_1,y) = 0$ for all $(x,y)$ in $S_1 \times S_2$.
\end{definition}

\begin{figure*}
\centering\includegraphics[width=0.45\textwidth]{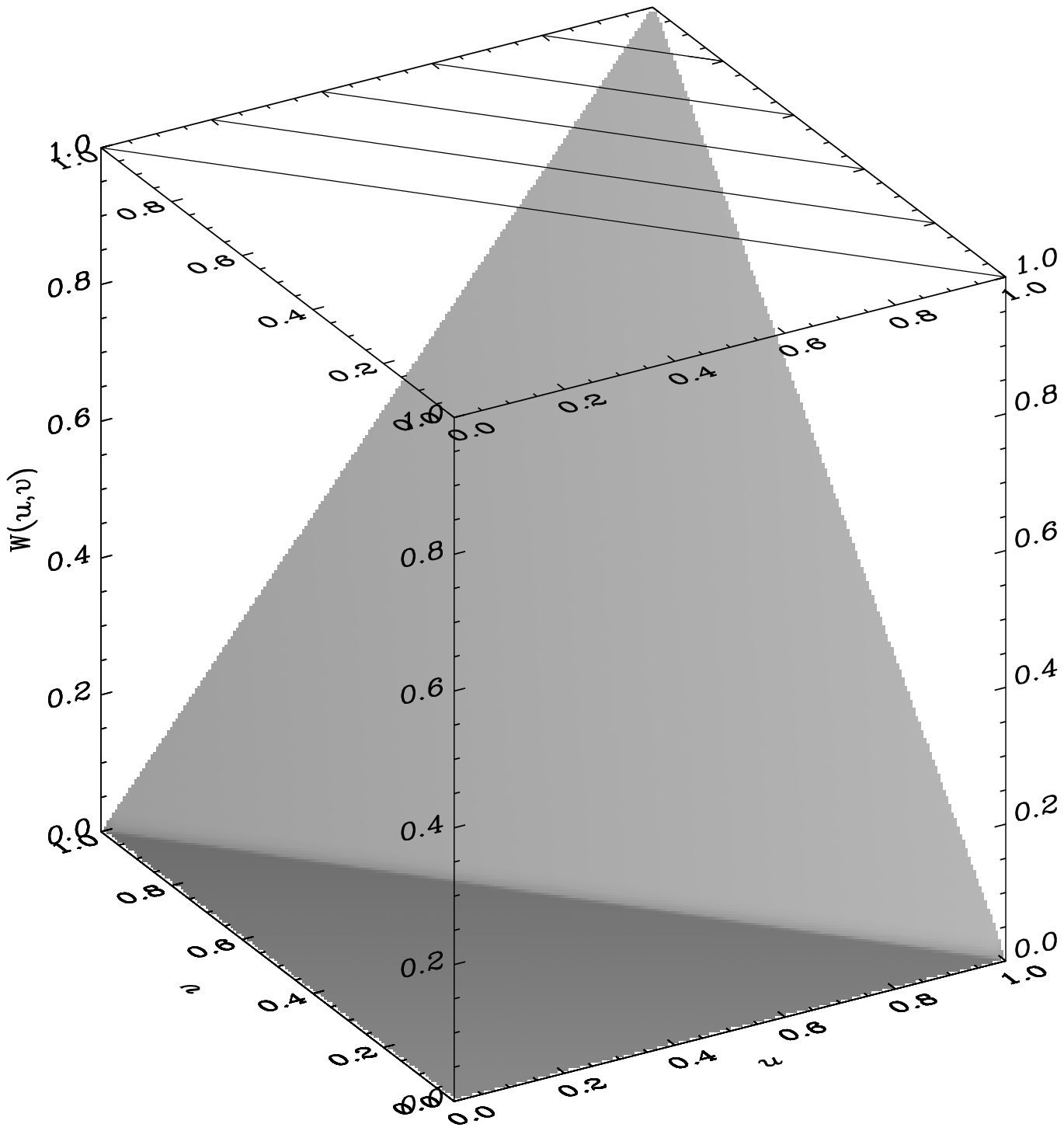}
\centering\includegraphics[width=0.45\textwidth]{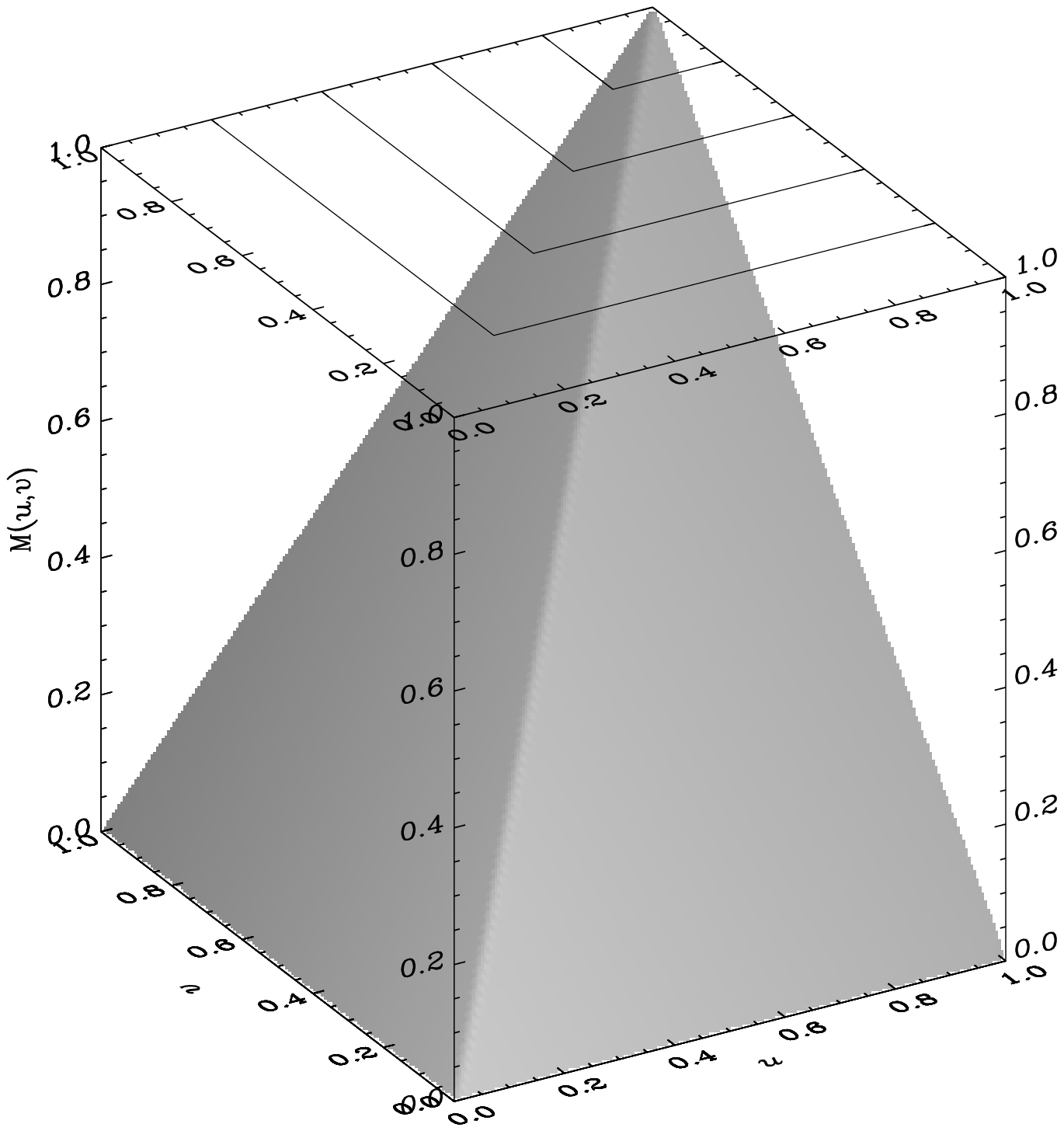}
\caption{
{ 
The Fr\'{e}chet-Hoeffding lower and upper bounds.
Left and right panels show the lower and upper bounds, respectively.
The top panels represent the contours of $W(u,v))$ and $M(u,v)$.
}
}\label{fig:FH}
\end{figure*}

With these concepts, we now define a two-dimensional copula.
\begin{definition}
  A two-dimensional copula (or shortly 2-copula) is a function $C$ with
  the following properties:
  \begin{enumerate}
  \item $\mbox{Dom } C = [0,1] \times [0,1]$;
  \item $C$ is grounded and 2-increasing; 
  \item For every $u$ and $v$ in [0,1], 
  \begin{eqnarray}
    C(u,1) = u \; \mbox{and} \; C(1,v) = v\;.
  \end{eqnarray}
  \end{enumerate}
\end{definition}
It may be useful to show there are upper and lower limits for the ranges of copulas, which
is given by the following theorem.
\begin{theorem}
  Let $C$ be a copula. 
  Then, for every $(u,v)$ in $\mbox{Dom } C$, 
  \begin{eqnarray}
     \max (u+v -1, 0) \leq C(u,v) \leq \min (u,v) \;.
  \end{eqnarray}
  Often the notations $W(u,v) \equiv \max(u+v -1,0)$ and $M(u,v) \equiv \min(u,v)$ are
  used.
  The former and the latter are referred to as {\it Fr\'{e}chet-Hoeffding lower bound} 
  and {\it Fr\'{e}chet-Hoeffding upper bound}, respectively.  
\end{theorem}
{
The Fr\'{e}chet-Hoeffding lower and upper bounds are illustrated in Fig.~\ref{fig:FH}. 
Any copula has its value between $W(u,v)$ and $M(u,v)$.
}

Any bivariate function which satisfies the above conditions can be a copula.
Then, there are infinite degrees of freedom for a set of copulas.
Using a copula $C$, we can construct a bivariate DF $G$ with two margins $F_1$ and $F_2$ as
\begin{eqnarray}
  G(x_1,x_2) = C[F(x_1),F_2(x_2)] \;.
\end{eqnarray}
However, one may have a natural question: can any bivariate DF be written in 
the above form? 
This is guaranteed by Sklar's theorem \citep{sklar59}.

\begin{theorem}{ (Sklar's theorem)}
Let $G$ be a joint distribution function with margins $F_1$ and $F_2$.
Then, there exists a copula $C$ such that for all $x_1, x_2$ in $\bar{\mathbb{R}}$,
\begin{eqnarray}\label{eq:copula}
  G(x_1,x_2) = C[F_1(x_1),F_2(x_2)] \;.
\end{eqnarray}
If $F_1$ and $F_2$ are continuous, then $C$ is unique: otherwise, $C$ is uniquely
determined on $\mbox{Range }F_1 \times \mbox{Range }F_2$.
\end{theorem}
A comprehensive proof Sklar's theorem is found in e.g., \citet{nelsen06}.
This theorem gives a basis that {\it any} bivariate DF with given margins is
expressed with a form of Equation~(\ref{eq:copula}).
Then, finally, the somewhat abstract definition of a copula turns out to be really 
useful for our aim, i.e., to construct a bivariate DF when its marginals are known in
some way.

Up to now, we discuss only bivariate DFs and their copulas. 
It is straightforward to introduce multivariate DFs as a natural extension of the formulation
presented here.

\subsection{Copulas and dependence measures between two variables}
\label{subsec:nonparam}

The most important statistical aspect of bivariate DFs is their
dependence properties between variables.
Since the dependence can never given by the marginals of a DF, this is the most
nontrivial information which a bivariate DF provides.
Since any bivariate DFs are described by Equation~(\ref{eq:copula}), all the information
on the dependence is carried by their copulas.

The most familiar measure of dependence among physical scientists (and others)
may be the correlation coefficients, especially Pearson's product-moment
correlation coefficient $\rho$.
The bivariate PDF of $x_1$ and $x_2$, $g(x_1,x_2)$, is written as
\begin{eqnarray}\label{eq:copula_density}
  g(x_1,x_2) = \frac{\partial^2 C[F_1(x_1),F_2(x_2)]}{\partial x_1 \partial x_2} f_1 (x_1) f_2 (x_2)
    \equiv c[F_1(x_1),F_2(x_2)] f_1(x_1) f_2(x_2)
\end{eqnarray}
where $f_1(x_1)$ and $f_2(x_2)$ are PDF of $F_1(x_1)$ and $F_2(x_2)$, respectively.
Then the correlation coefficient is expressed as
\begin{eqnarray}\label{eq:corr}
  \rho = \frac{\int (x_1 - \bar{x_1})(x_2 - \bar{x_2}) g(x_1,x_2) 
    \pd x_1 \pd x_2}{\sqrt{\int (x_1 - \bar{x_1})^2 f_1(x_1) \pd x_1 
    \int (x_2 - \bar{x_2})^2 f_2(x_2) \pd x_2}}  = 
  \frac{\int (x_1 - \bar{x_1})(x_2 - \bar{x_2})  c[(F_1(x_1),F_2(x_2)] f_1(x_1) f_2(x_2) \pd x_1 \pd x_2}{
    \sqrt{\int (x_1 - \bar{x_1})^2 f_1(x_1) \pd x_1 \int (x_2 - \bar{x_2})^2 f_2(x_2) \pd x_2}} \;.
\end{eqnarray}
We should note that $\rho$ can measure only {\it a linear dependence} of two variables.
However, in general the dependence of two variables would be not linear at all, and it cannot be 
a sufficient measure of dependence.
Further, more fundamentally, Equation~(\ref{eq:corr}) depends not only on the dependence of two
variables (copula part) but also its marginals $f_1(x_1), f_2(x_2)$, i.e.,
the linear correlation coefficient $\rho$ does not measure the dependence purely.
In such a situation, a more flexible and genuine measure of dependence, e.g., Spearman's $\rs$
or Kendall's $\tau$ would be more appropriate.
Spearman's rank correlation is a nonparametric version of Pearson's correlation using ranked data.
The population version of Spearman's $\rs$ is expressed by copula as
\begin{eqnarray}\label{eq:spearman_copula}
  \rs \hspace{-3mm}&=&\hspace{-3mm} 
    12 \int_0^1 \int_0^1 u_1 u_2 \pd C(u_1,u_2) - 3 \nonumber \\ 
  \hspace{-3mm}&=&\hspace{-3mm} 
    12 \int_0^1 \int_0^1 C(u_1, u_2) \pd u_1 \pd u_2 -3 \;. 
\end{eqnarray}
The definition of Kendall's tau is more complicated. 
We define a concept of concordance as follows:
when we have pairs of data $(x_{1i},x_{2i})$ and $(x_{1j},x_{2j})$, they are said to be
{\it concordant} if $x_{1i} > x_{1j}$ and $x_{2i} > x_{2j}$ or $x_{1i} < x_{1j}$ and $x_{2i} < x_{2j}$ 
(i.e., $(x_{1i} - x_{1j})(x_{2i} - x_{2j}) > 0$), and otherwise {\it discordant}.
Let $\{x_{1i},x_{2i}\}_{i=1,\dots,n}$ denote a random sample of $n$ observations.
There are ${}_nC_2$ pairs $(x_{1i}, x_{2i})$ and $(x_{1j}, x_{2j})$ of observations in
the sample, and each pair is concordant or discordant. 
Let $\nc$ denote the number of concordant pairs, and $\nd$ the number of discordant
ones.
Then, Kendall's $\tau$ for the sample, $t$, is defined as
\begin{eqnarray}\label{eq:kendall_sample}
  t = \frac{\nc - \nd}{\nc + \nd} \;.
\end{eqnarray} 
The population version of $\tau$ is also expressed in a simple form in terms of copula as
\begin{eqnarray}\label{eq:kendall_copula}
    \tau \hspace{-3mm}&=&\hspace{-3mm} 
    4 \int_0^1 \int_0^1 C(u_1,u_2) \pd C(u_1,u_2) - 1 \nonumber \\ 
  \hspace{-3mm}&=&\hspace{-3mm} 
    4 \int_0^1 \int_0^1 C(u_1, u_2) c(u_1,u_2) \pd u_1 \pd u_2 -1 \;. 
\end{eqnarray}
Note that both Equations~(\ref{eq:spearman_copula}) and (\ref{eq:kendall_copula}) are 
independent of the distributions $F_1, F_2$ and $G$, but they only show the
dependence structure described by the copula, unlike the linear correlation
coefficient.
This is a direct consequence of the non-parametric (i.e., distribution-free) nature 
of these estimators. 
These are the reasons why both dependence measures are almost always 
used in the context of copulas in the literature.
Estimators of $\rs$ and $\tau$ from a sample are found in Appendix~\ref{sec:nonparam_est}.

\subsection{Farlie--Gumbel--Morgenstern (FGM) copula}
\label{subsec:fgm}

As seen in the discussion above, usefulness of the linear correlation coefficient is
quite limited, and distribution-free measures of dependence are more appropriate 
for general joint DFs with non-Gaussian marginals.
However, even if it is true, since physicists may cling to the most familiar linear
correlation coefficient $\rho$, a copula which has an explicit dependence on $\rho$ 
would be convenient.
We introduce two special types of copulas with this ideal property.

For cases where the correlation between two variables is weak,
a systematic method has been proposed by \citet{morgenstern56} and 
\citet{gumbel60} for specific functional forms, and later generalized to 
arbitrary functions by \citet{farlie60}.
This is known as the Farlie--Gumbel--Morgenstern (FGM) distributions after the
inventors' names.
Although the study of the FGM distributions does not seem tightly connected to 
that of copulas, we will see later that they can be expressed in terms of the 
so-called FGM copula.

The correlation structure of the FGM distributions was studied by
\citet{schucany78}.
Let $F_1(x_1)$ and $F_2(x_2)$ be the (cumulative) distribution functions (DFs) 
of a stochastic variables $x$ and $y$, respectively, 
and let $f_1(x_1)$ and $f_2(x_2)$ be their probability density functions 
(PDFs).
The bivariate FGM system of distributions $G(x_1,x_2)$
is written as
\begin{eqnarray}\label{eq:cum_fgm}
  G(x_1,x_2) = F_1(x_1)F_2(x_2) \left\{ 1+\kappa \left[1-F_1(x_1)\right]
    \left[1-F_2(x_2)\right]\right\} \,
\end{eqnarray}
where $G(x_1,x_2)$ is the joint DF of $x_1$ and $x_2$.
Here $\kappa$ is a parameter related to the correlation (see below), and to 
make $G(x_1,x_2)$ have an appropriate property as a bivariate DF, 
$\left| \kappa \right| \le 1$ is required 
\citep[for a proof, see][]{cambanis77}.
Its PDF can be obtained by a direct differentiation of $G(x_1,x_2)$ as
\begin{eqnarray}\label{eq:dif_fgm}
  g(x_1,x_2) \hspace{-3mm}&\equiv & \hspace{-3mm}
    \left. \frac{\partial^2 G}{\partial x_1 \partial x_2} \right|_{x_1,x_2} 
    = \left. \frac{\partial^2}{\partial x_1 \partial x_2}\left \{
    F_1(x_1)F_2(x_2) \left[ 1+\kappa \left(1-F_1(x_1)\right)
    \left(1-F_2(x_2)\right)\right] \right\} \right|_{x_1,x_2}\nonumber \\
  \hspace{-3mm}&=&\hspace{-3mm} 
    f_1(x_1) f_2(x_2) \left\{ 1+\kappa \left[2F_1(x_1)-1\right]
    \left[2F_2(x_2)-1\right]\right\}\;.
\end{eqnarray}
{}From Equation~(\ref{eq:dif_fgm}), it is straightforward to obtain its
covariance function $\mathsf{Cov}(x_1,x_2)$ as
\begin{eqnarray}\label{eq:cov_fgm}
  \mathsf{Cov}(x_1,x_2) =
     \kappa \int x_1 \left[2F_1(x_1)-1 \right] f_1(x_1) \pd x_1 
     \int x_2 \left[2F_2(x_2)-1 \right] f_2(x_2) \pd x_2 \;.
\end{eqnarray}
Then we have a correlation function of two stochastic variables $x_1$ and 
$x_2$, $\rho(x_1,x_2)$ [Equation~(\ref{eq:corr})] as follows
\begin{eqnarray}
  \rho(x_1,x_2) = \frac{\mathsf{Cov}(x_1,x_2)}{\sigma_1 \sigma_2}
    =\frac{\kappa}{\sigma_1 \sigma_2} \int x_1 \left[2F_1(x_1)-1 \right] f_1(x_1) \pd x_1 
     \int x_2 \left[2F_2(x_2)-1 \right] f_2(x_2) \pd x_2 \;.
\end{eqnarray}
where $\sigma_1$ and $\sigma_2$ are the standard deviations of $x_1$ and $x_2$ 
with respect to $f_1(x_1)$ and $f_2(x_2)$.
It is straightforwardly confirmed that $g(x_1,x_2)$ really has the marginals
$f_1(x_1)$ and $f_2(x_2)$, by a direct integration with respect to $x_1$ 
or $x_2$.
It is also clear that if we want a bivariate PDF with a prescribed 
correlation coefficient $\rho$,
we can determine the parameter $\kappa$ from Equations~(\ref{eq:corr})
and (\ref{eq:cov_fgm}).

Here, consider the case that that both $f_1(x_1)$ and $f_2(x_2)$ are the Gamma 
distributions, i.e.,
\begin{eqnarray}
  f_j(x_j)= \frac{{x_j}^{a-1} e^{-{x_j}/b}}{b^a \Gamma (a)}
\end{eqnarray}
where $\Gamma(x)$ is the gamma function ($j=1,2$).
In this case, after some algebra, $\rho$ can be written analytically as
\begin{eqnarray}
  \rho(x_1,x_2) = \frac{\kappa}{\sqrt{ab}} 
    \left[ \frac{2^{-2 (a -1)}}{B(a,a)}\right]
    \left[ \frac{2^{-2 (b -1)}}{B(b,b)}\right] \;
\end{eqnarray}
where
\begin{eqnarray}
  B(a,b) \equiv \frac{\Gamma (a)\Gamma (b)}{\Gamma (a+b)} \;.
\end{eqnarray}
This result was obtained by \citet{deste81}.
Especially when $b=1$, this corresponds to a bivariate extension of the 
Schechter function \citep{schechter76}, and we expect some astrophysical
applications.

As we mentioned, the correlation of the FGM distributions is restricted to be 
weak: indeed, the correlation coefficient cannot exceed $1/3$.
Here we prove this.
For all (absolutely continuous) $F(x)$,
\begin{eqnarray}
  \left\{\int x \left[2F(x)-1 \right] f(x) \pd x\right\}^2 
    \hspace{-3mm}&=&\hspace{-3mm}
    \left\{\int \left(x-\bar{x}\right) \left[2F(x)-1 \right] f(x) \pd x\right\}^2 
    \nonumber \\
    \hspace{-3mm}&\le&\hspace{-3mm}
    \int \left(x-\bar{x}\right)^2  f(x) \pd x
    \int \left[2F(x)-1 \right]^2 f(x) \pd x
    =\frac{\sigma^2}{3} \;,
\end{eqnarray}
where $\bar{x}$ is the average of $x$.
The second line follows from the Cauchy--Schwarz inequality.
{}From Equations~(\ref{eq:cov_fgm}) and (\ref{eq:corr}), and the
condition $\left| \kappa \right| \le 1$, 
we obtain $\left| \rho \right| \le 1/3$.

The copula of the FGM family of distributions is expressed as
\begin{eqnarray}\label{eq:fgm_copula}
  C^{\rm FGM} (u_1, u_2; \kappa) = u_1 u_2 + \kappa u_1 u_2 (1-u_1) (1-u_2)  
\end{eqnarray}
with $-1 < \kappa < 1$.
The differential form of the FGM copula follows from Equations~(\ref{eq:copula_density}) and (\ref{eq:fgm_copula})
\begin{eqnarray}\label{eq:fgm_copula_density}
  c^{\rm FGM} (u_1, u_2; \kappa) = 1+ \kappa (1-2u_1) (1-2u_2) \; . 
\end{eqnarray}

\subsection{Gaussian copula}
\label{subsec:copula_gauss}

As seen in Section~\ref{subsec:fgm}, though the FGM distribution has one of the most ``natural'' 
example of bivariate DF which has an explicit $\rho$-dependence,
the limitation of the correlation coefficient of the FGM family hampers a flexible application of 
this DF, though there have been many attempts to extend its range of application 
(see Appendix~\ref{sec:jk77}).
Then, the second natural candidate may be a copula related to a bivariate Gaussian DF.
The Gaussian copula has also an explicit dependence on a linear correlation
coefficient by its construction.

Let
\begin{eqnarray}
  &&\psi_1 (x) = \frac{1}{\sqrt{2\pi}} \exp\left(-\frac{x^2}{2}\right) \;, \label{eq:1gauss}\\
  &&\Psi_1 = \int_{-\infty}^{x} \Psi (x') \pd x' \;, \\
  &&\psi_2 (x_1, x_2; \rho) = \frac{1}{\sqrt{(2\pi)^2(1-\rho^2)}} 
    \exp\left[-{\frac{x_1^2+x_2^2-2\rho x_1x_2}{2(1-\rho^2)}}\right] \;, \label{eq:2gauss}
\end{eqnarray}
and
\begin{eqnarray}
  \Psi_2 (x_1, x_2; \rho) = \int_{-\infty}^{x_1} \int_{-\infty}^{x_2} \psi_i (x_1', x_2') \pd x_1' \pd x_2' \;.
\end{eqnarray}
By using the covariance matrix $\vmatrix$
\begin{eqnarray}\label{eq:vmatrix}
  \vmatrix \equiv  
    \begin{pmatrix}
      1 & \rho \\
      \rho & 1 
    \end{pmatrix} \;,
\end{eqnarray}
Equation~(\ref{eq:2gauss}) is simplified as
\begin{eqnarray}
  \psi_2 (x_1, x_2; \rho) = \frac{1}{\sqrt{(2\pi)^2 \det \vmatrix}} 
    \exp\left(-\frac{1}{2} \vx^T \vmatrix^{-1} \vx \right) \;, \label{eq:2gauss_matrix} 
\end{eqnarray}
where $\vx \equiv (x_1, x_2)^T$ and superscript $T$ stands for the transpose of 
a matrix or vector.

Then, we define a Gaussian copula $C^{\rm G}(u_1, u_2; \rho)$ as
\begin{eqnarray}
  C^{\rm G} (u_1, u_2; \rho) = \Psi_2 \left[ \Psi_1^{-1} (u_1), \Psi_1^{-1} (u_2) ; \rho \right] \;.
\end{eqnarray}
The density of $C^{\rm G}$, $c^G$, is obtained as
\begin{eqnarray}
  c^{\rm G} (u_1, u_2; \rho) \hspace{-3mm}&=&\hspace{-3mm} 
    \frac{\partial^2 C^{\rm G} (u_1, u_2; \rho)}{\partial u_1 \partial u_2} 
    =\frac{\partial^2 \Psi_2 \left[ \Psi_1^{-1} (u_1), \Psi_1^{-1} (u_2) ; \rho \right]}{
    \partial u_1 \partial u_2} \nonumber \\
  \hspace{-3mm}&=&\hspace{-3mm} 
    \frac{\psi_2 (x_1, x_2; \rho)}{\psi_1 (x_1) \psi_1 (x_2)} \nonumber \\
  \hspace{-3mm}&=&\hspace{-3mm} 
    \frac{\mathstrut \displaystyle
    \frac{1}{\sqrt{(2\pi)^2 \det \vmatrix}}\exp\left(-\frac{1}{2} \vx^T \vmatrix^{-1} \vx \right)}{
    \mathstrut \displaystyle
    \frac{1}{\sqrt{2\pi}} \exp\left(-\frac{x_1^2}{2}\right) \frac{1}{\sqrt{2\pi}} \exp\left(-\frac{x_2^2}{2}\right)}
    \nonumber \\
  \hspace{-3mm}&=&\hspace{-3mm} 
    \frac{1}{\sqrt{\det \vmatrix}}
    \exp\left[-\frac{1}{2} \left( \vx^T \vmatrix^{-1} \vx - \vx^T \imatrix \vx \right) \right]  
    \nonumber \\
  \hspace{-3mm}&=&\hspace{-3mm} 
    \frac{1}{\sqrt{\det \vmatrix}}
    \exp\left\{-\frac{1}{2} \left[ \vx^T \left(\vmatrix^{-1} - \imatrix \right) \vx \right] \right\} 
    \nonumber \\
  \hspace{-3mm}&=&\hspace{-3mm} 
    \frac{1}{\sqrt{\det \vmatrix}}
    \exp\left\{-\frac{1}{2} \left[ \vpsi^T \left(\vmatrix^{-1} - \imatrix \right) \vpsi \right] \right\} \;,  
\end{eqnarray}
where $\vpsi \equiv \left[ \Psi^{-1}(u_1), \Psi^{-1}(u_2) \right]^T$ and $\imatrix$ stands for the identity
matrix.
The second line follows from Equation~(\ref{eq:copula_density}).

\begin{figure*}
\centering\includegraphics[angle=0,width=0.3\textwidth]{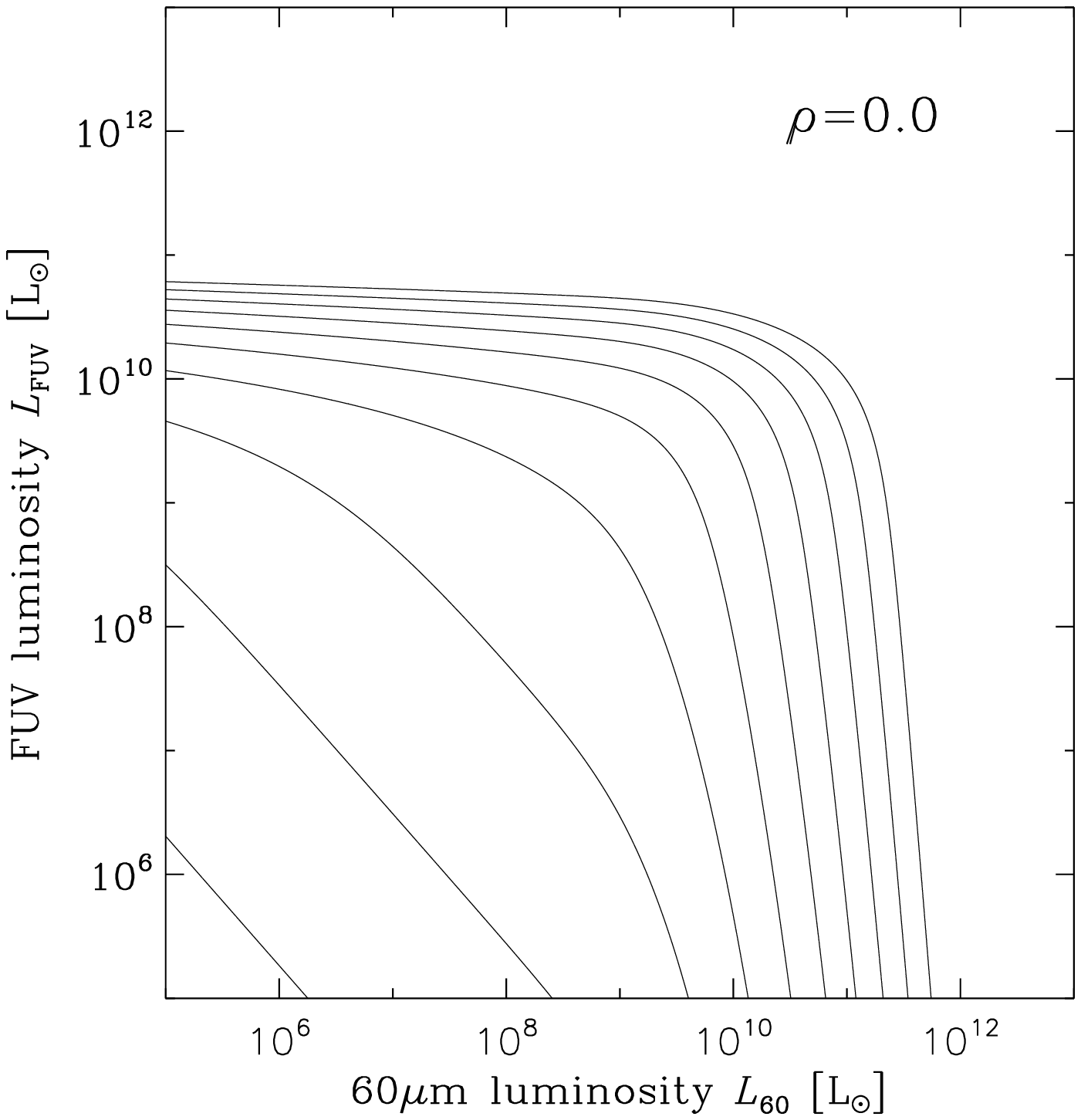}
\centering\includegraphics[angle=0,width=0.3\textwidth]{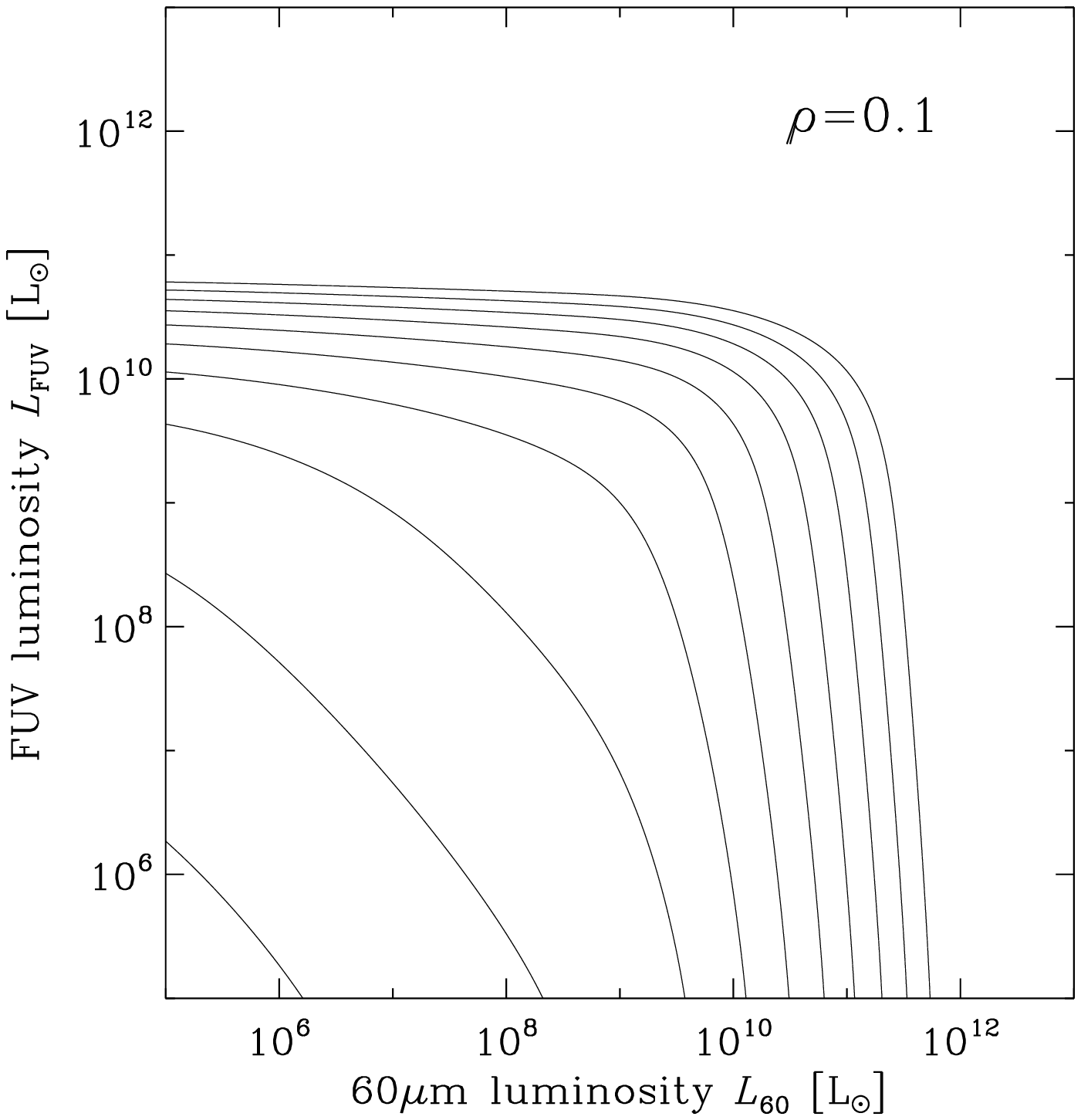}\\
\noindent
\centering\includegraphics[angle=0,width=0.3\textwidth]{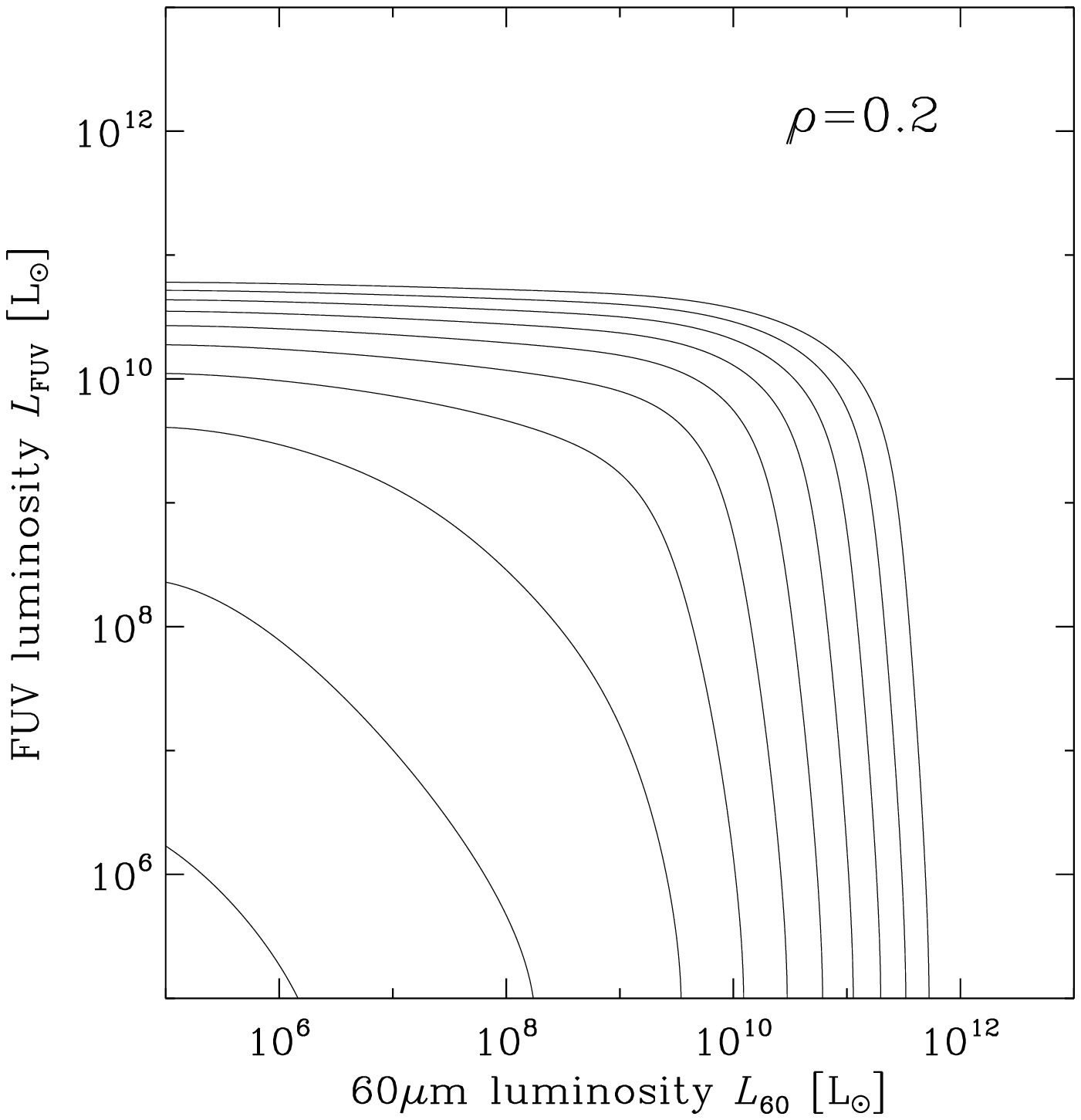}
\centering\includegraphics[angle=0,width=0.3\textwidth]{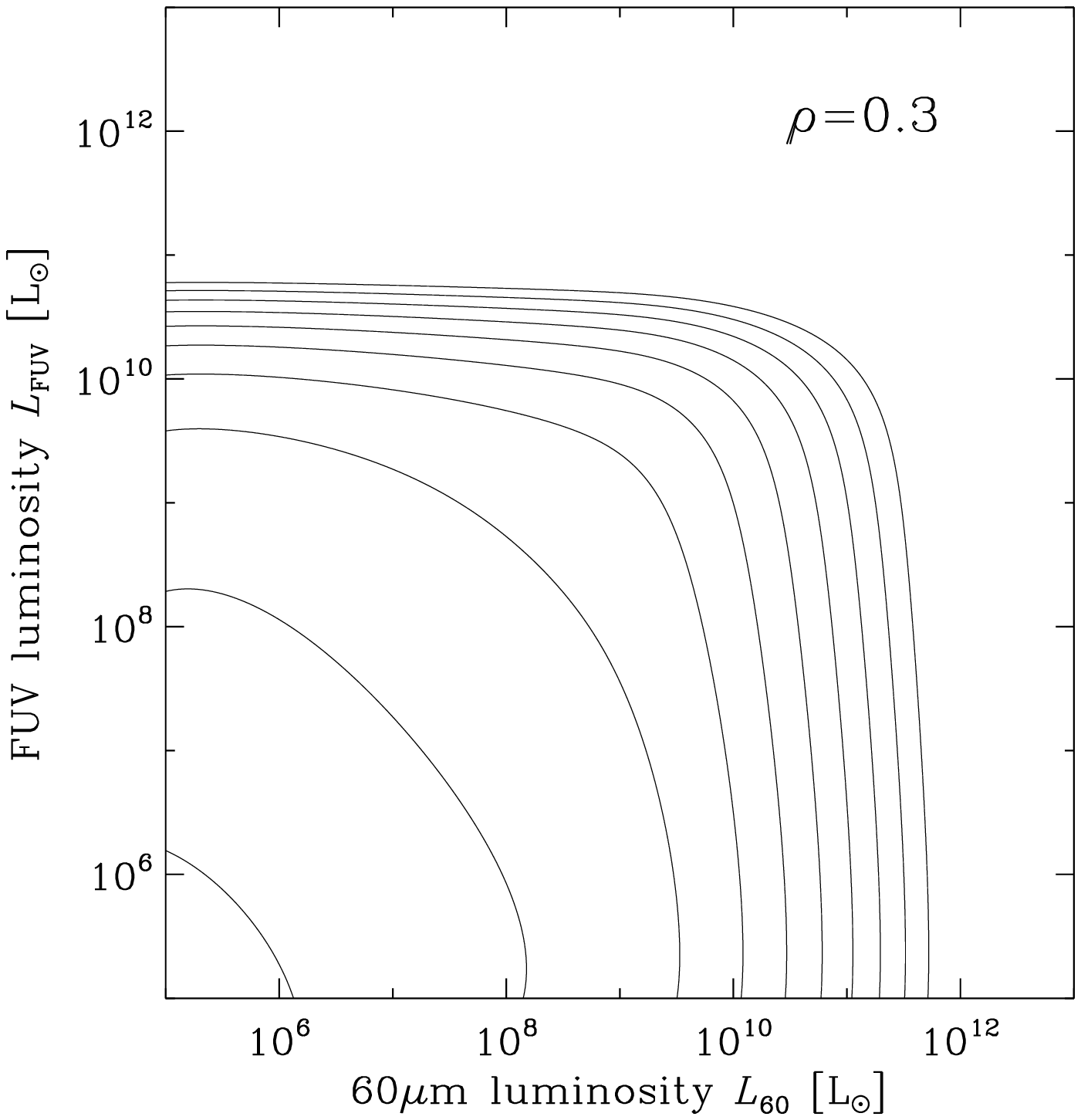}
\caption{
{
The bivariate luminosity functions (BLFs) constructed with
the Farlie--Gumbel--Morgenstern copula, with model luminosity functions
(LFs) of ultraviolet (UV) and infrared (IR)-selected galaxies.
The BLFs are normalized so that integrating it over the whole ranges of 
$L_1$ and $L_2$ gives one.
}
{}From top-left to bottom-right, the linear correlation coefficient $\rho=0.0$, 0.1, 0.2, and 0.3, 
corresponding to $\kappa=0.0, 0.33, 0.67$ and 1.0, respectively.
{
The contours are logarithmic with an interval $\Delta \log \phi^{(2)} = 0.5$ drawn from the
peak probability. 
}
}
\label{fig:fgm_lf}
\end{figure*}

\section{Application to construct the bivariate luminosity function (BLF) of galaxies}
\label{sec:blf}

\subsection{Construction of the BLF}

We define the luminosity at a certain wavelength band by 
$L\equiv\nu L_\nu$ ($\nu$ is the corresponding frequency).
Then the luminosity function is defined as a number density of galaxies whose
luminosity lies between a logarithmic interval
$[\log L, \log L + \pd\log L]$: 
\begin{eqnarray}
 \phi^{(1)} (L) \equiv \frac{\pd n}{\pd \log L} \;,
\end{eqnarray}
where we denote $\log x \equiv \log_{10} x$ and $\ln x \equiv \log_e x$.
For mathematical simplicity, we define the LF as {\sl being normalized}, i.e., 
\begin{eqnarray}
 \int \phi^{(1)} (L) \pd\log L = 1\;.
\end{eqnarray}
Hence, this corresponds to a probability density function (PDF), a commonly
used terminology in the field of mathematical statistics.
We also define the cumulative LF as
\begin{eqnarray}
 \Phi^{(1)} (L) \equiv \int_{\log L_{\rm min}}^{\log L} \phi^{(1)} (L') \pd \log L' \;,
\end{eqnarray}
where $L_{\rm min}$ is the minimum luminosity of galaxies considered.
This corresponds to the DF.

\begin{figure*}
\centering\includegraphics[angle=0,width=0.3\textwidth]{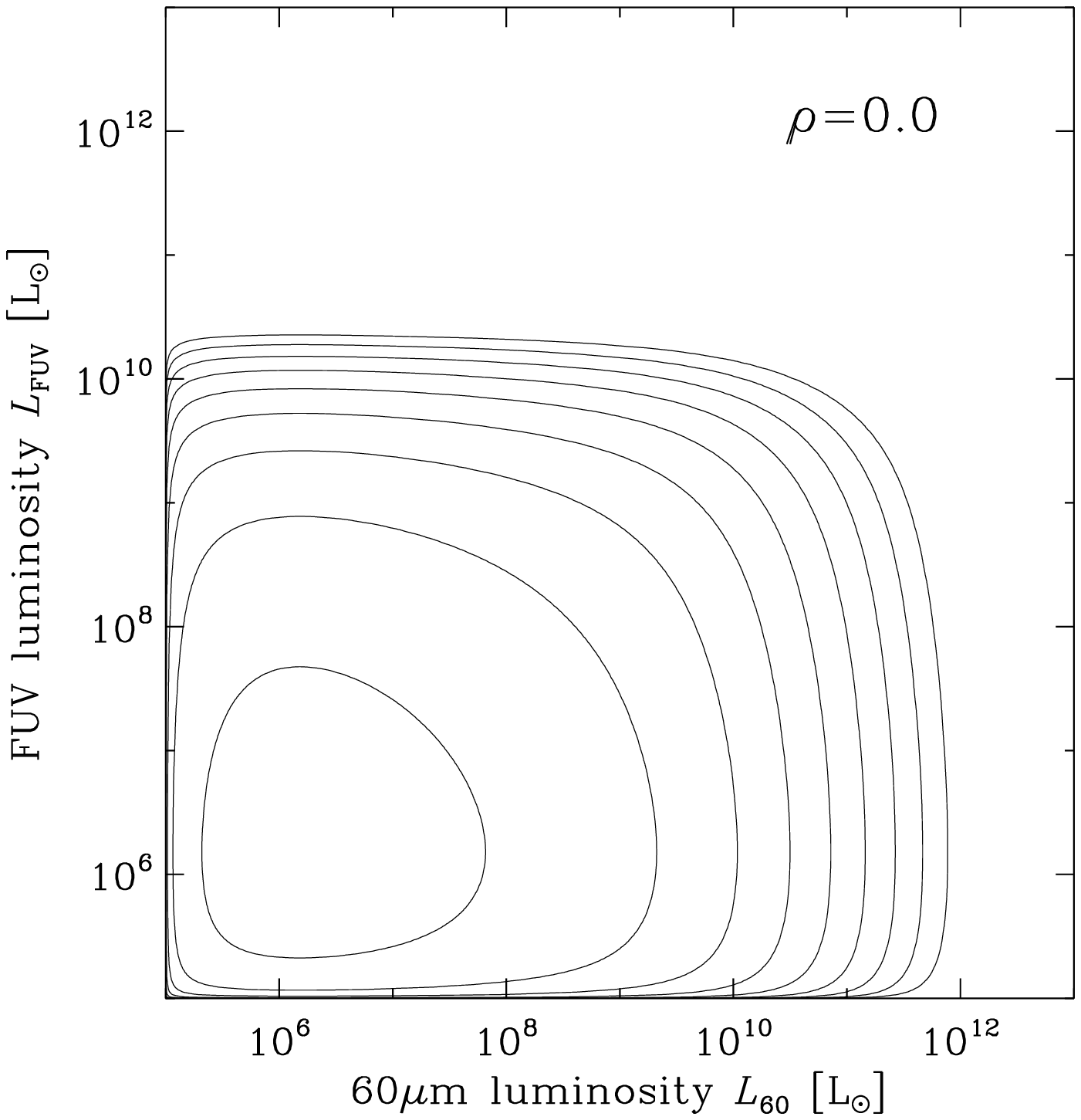}
\centering\includegraphics[angle=0,width=0.3\textwidth]{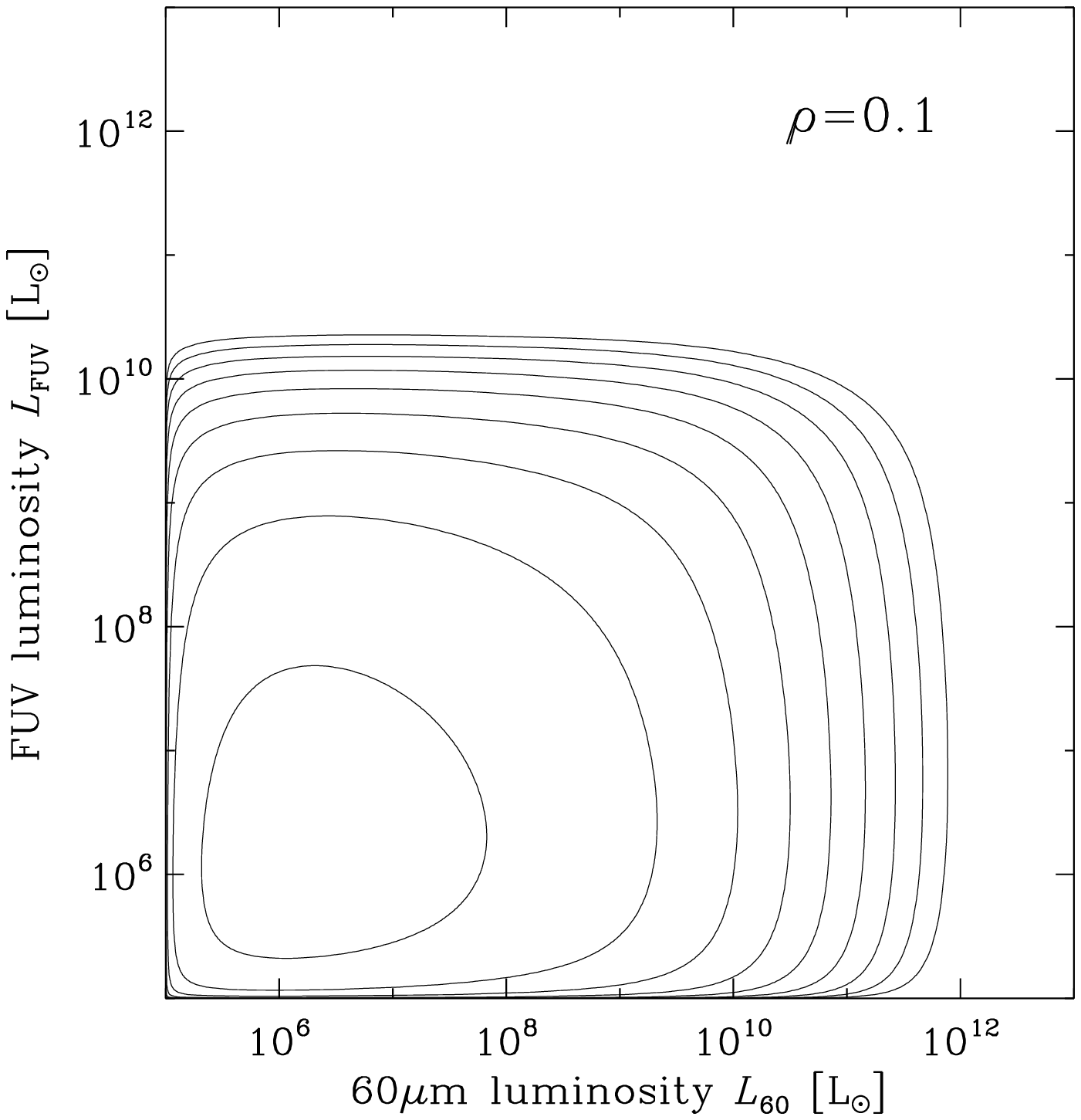}
\centering\includegraphics[angle=0,width=0.3\textwidth]{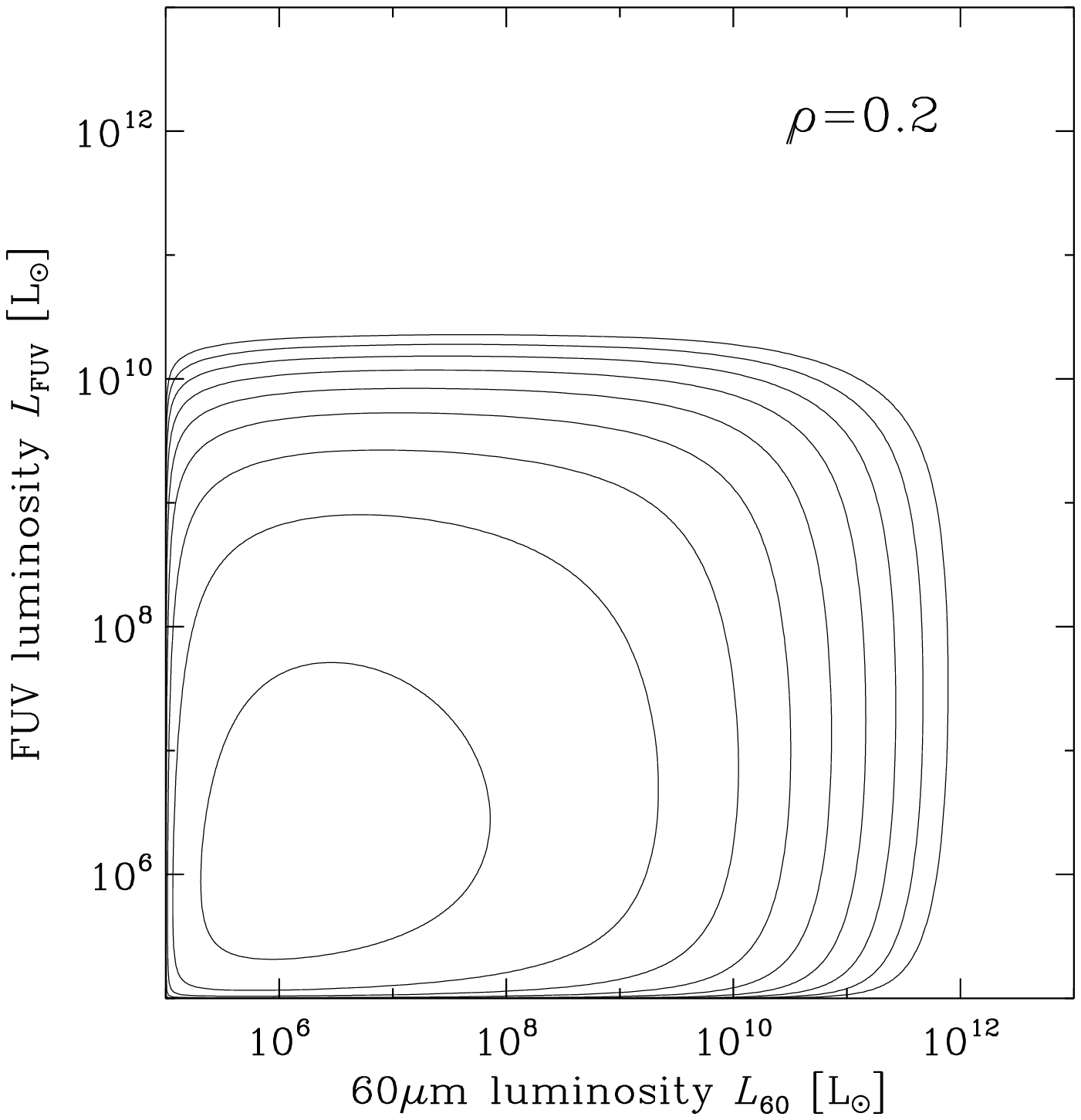}
\centering\includegraphics[angle=0,width=0.3\textwidth]{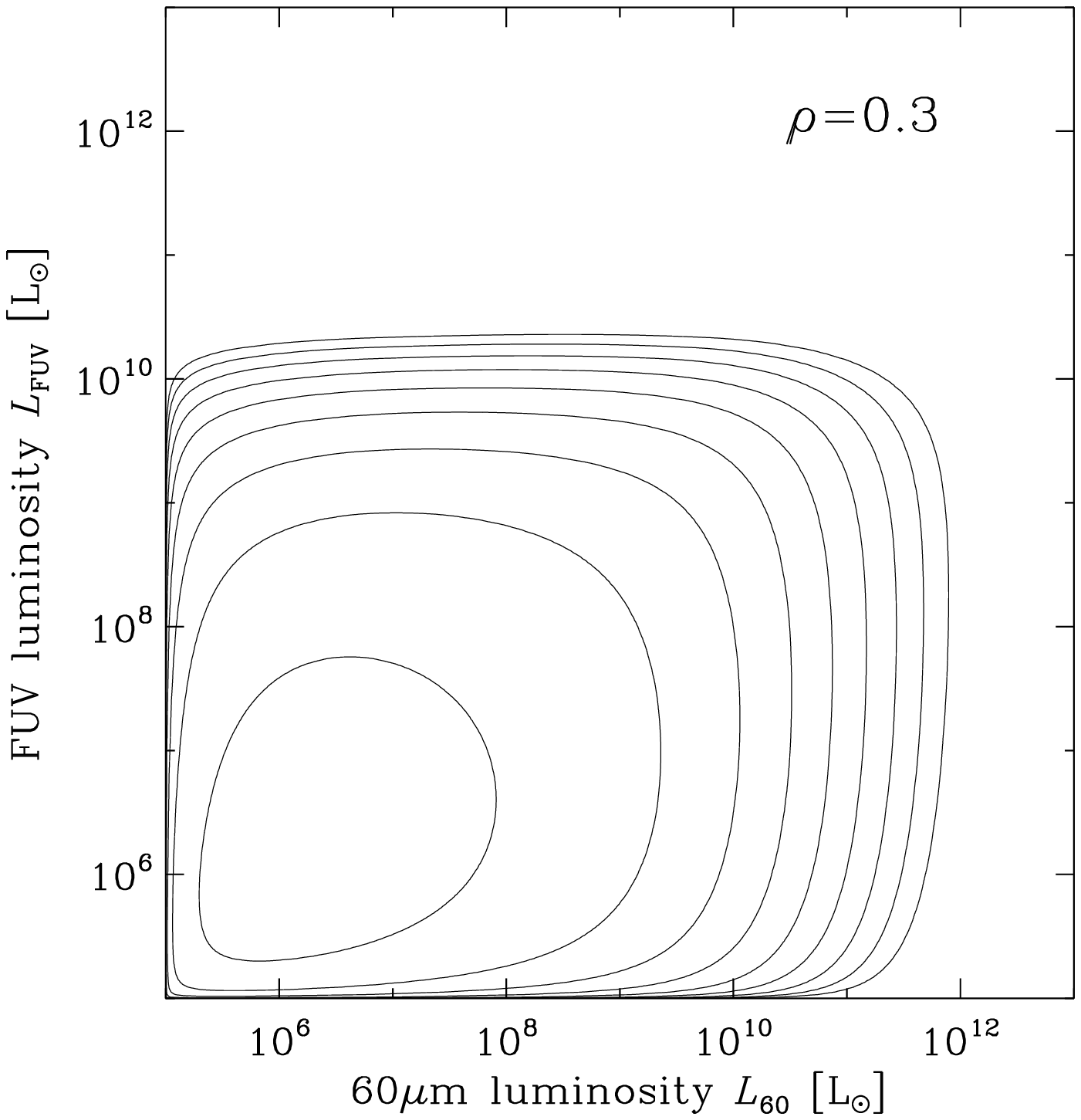}
\centering\includegraphics[angle=0,width=0.3\textwidth]{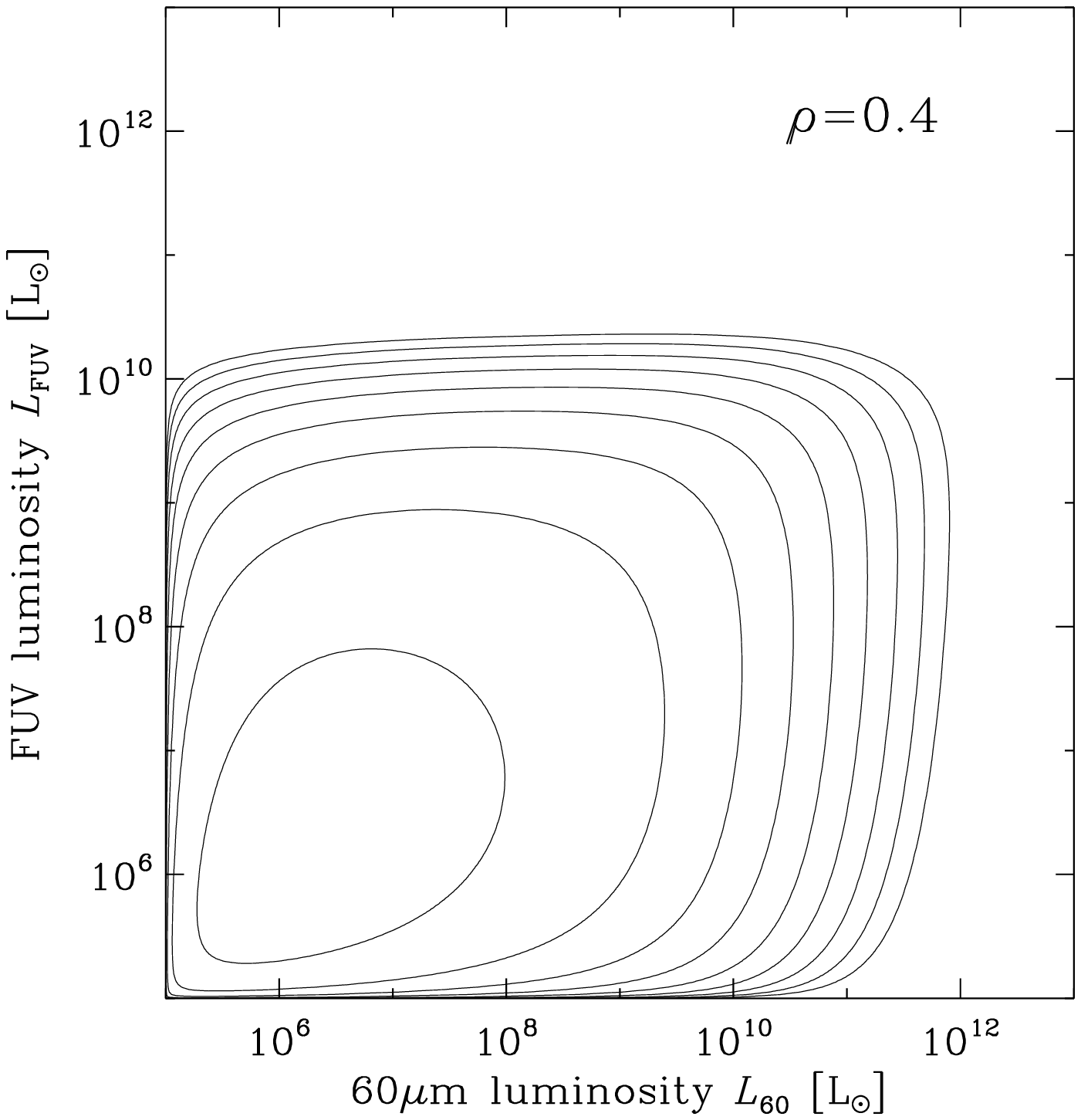}
\centering\includegraphics[angle=0,width=0.3\textwidth]{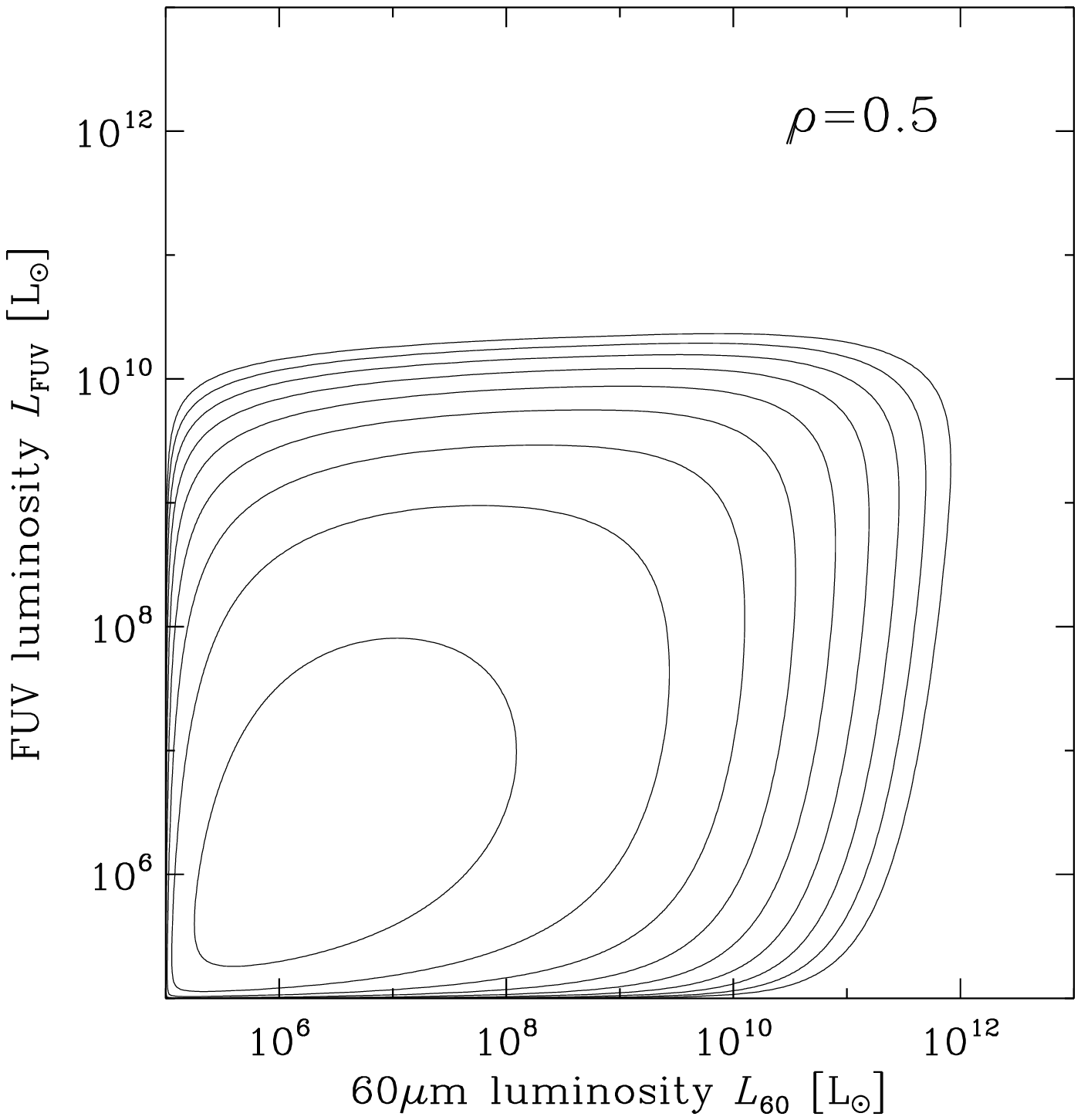}
\centering\includegraphics[angle=0,width=0.3\textwidth]{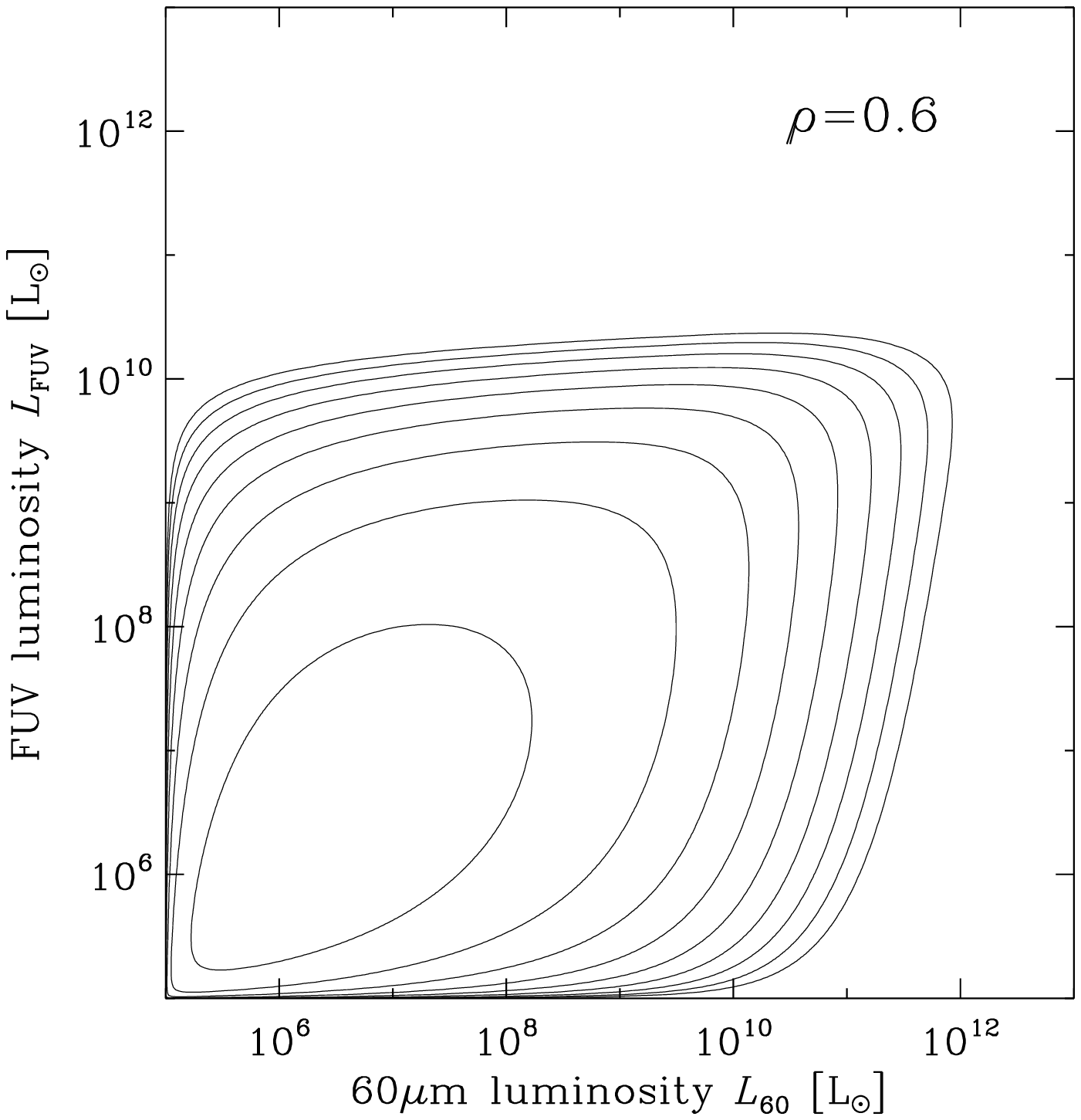}
\centering\includegraphics[angle=0,width=0.3\textwidth]{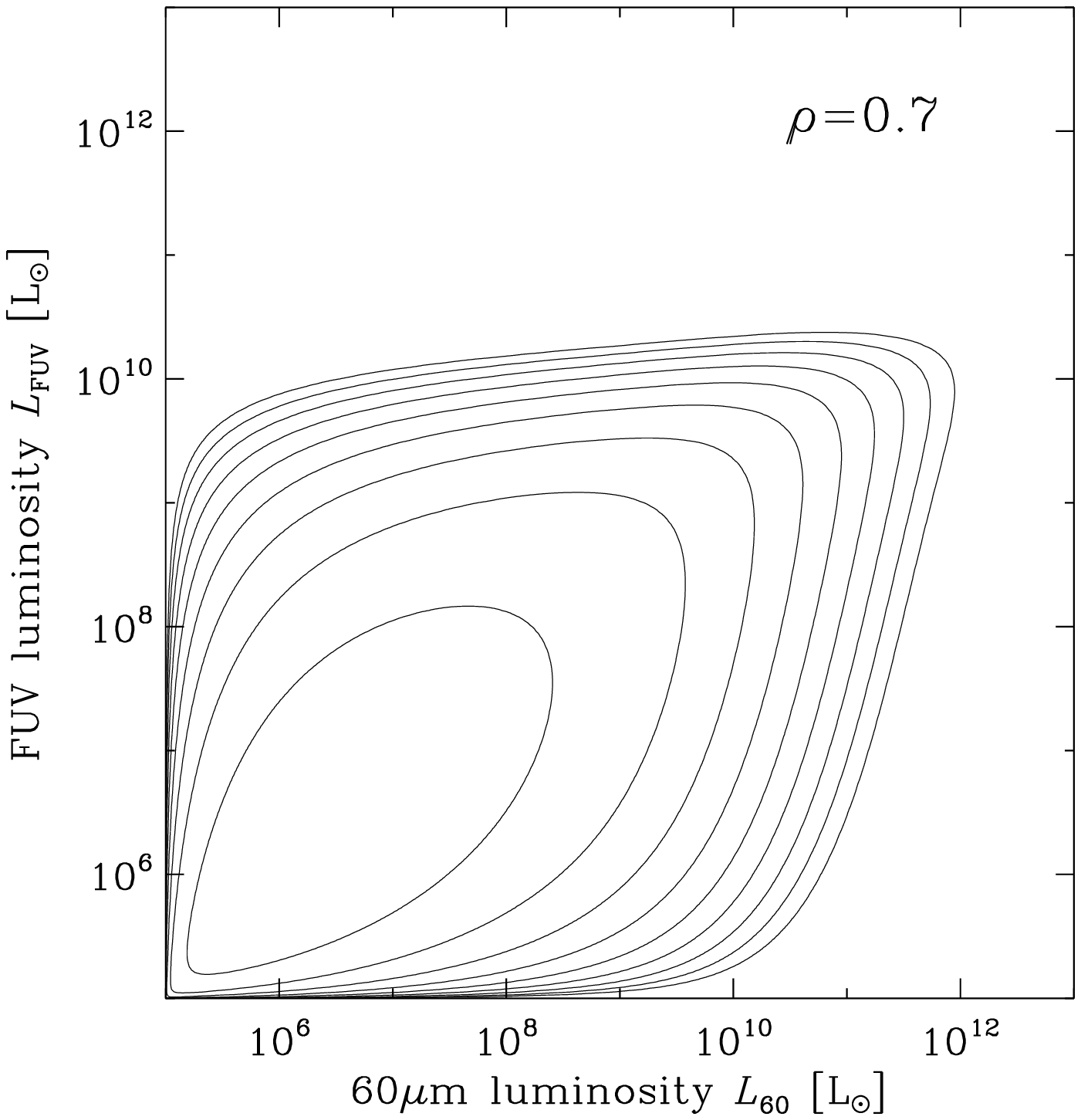}
\centering\includegraphics[angle=0,width=0.3\textwidth]{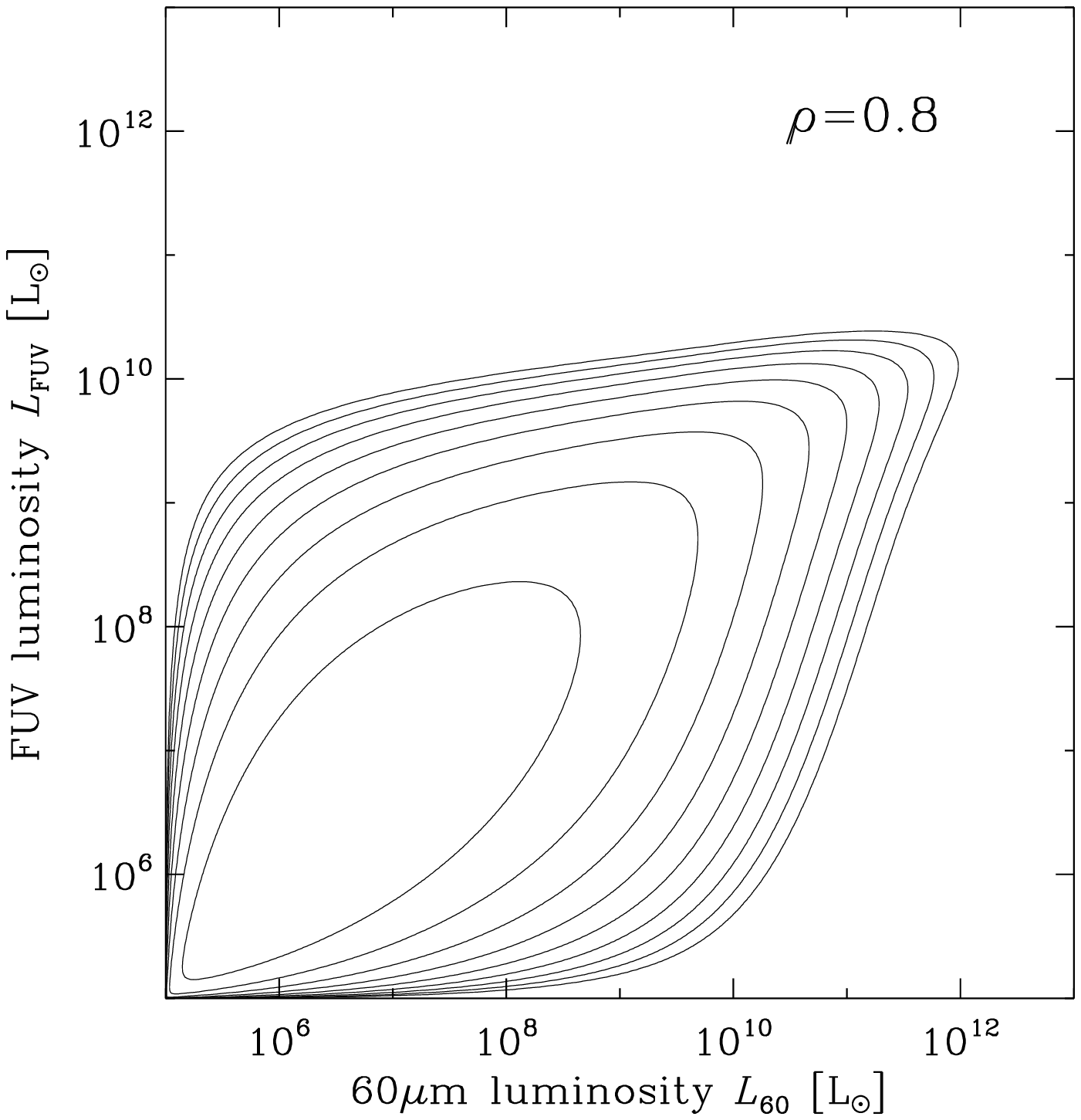}
\centering\includegraphics[angle=0,width=0.3\textwidth]{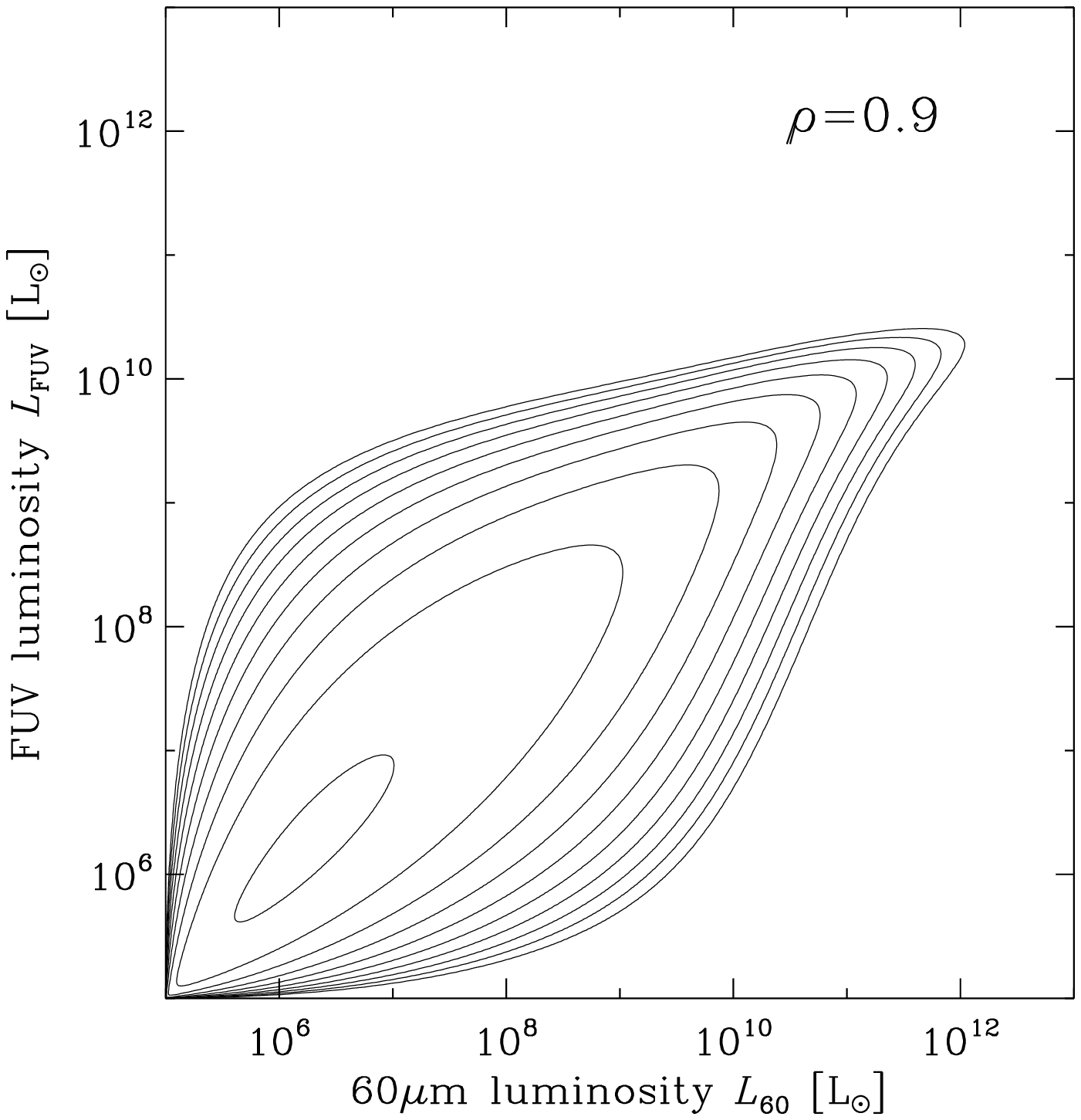}
\caption{
{
The analytical BLF constructed with the Gaussian copula with model 
luminosity functions (LFs) of ultraviolet (UV) and infrared (IR)-selected galaxies..
The BLFs are again normalized so that integrating it over the whole ranges of 
$L_1$ and $L_2$ gives one.
The linear correlation coefficient $\rho$ varies from 0.0 to 0.9 from top-left to
bottom-right.
The same as Fig.~\ref{fig:fgm_lf}, the contours are logarithmic with 
an interval $\Delta \log \phi^{(2)} = 0.5$ drawn from the peak probability. 
}
}
\label{fig:gauss_lf}
\end{figure*}

If we denote univariate LFs as $\phi^{(1)}_1(L_1)$ and $\phi^{(1)}_2(L_2)$, then
the bivariate PDF $\phi^{(2)}(L_1,L_2)$ is described by a differential copula $c(u_1, u_2)$ as
\begin{eqnarray}
  \phi^{(2)} (L_1, L_2) \hspace{-3mm}&\equiv & \hspace{-3mm}
    c \left[\phi^{(1)}_1(L_1), \phi^{(1)}_2(L_2)\right] \;.
\end{eqnarray}
For the FGM copula, the BLF leads from Equation~(\ref{eq:fgm_copula_density})
\begin{eqnarray}\label{eq:lf_fgm}
  \phi^{(2)}(L_1,L_2 ; \kappa) \hspace{-3mm}&\equiv&\hspace{-3mm}
    \left\{ 1+\kappa \left[2\Phi^{(1)}_1(L_1)-1\right]
    \left[2\Phi^{(1)}_2(L_2) - 1 \right] \right\} \phi^{(1)}_1 (L_1) \phi^{(1)}_2 (L_2) \; .
\end{eqnarray}
The parameter $\kappa$ is proportional to the correlation coefficient
$\rho$ between $\log L_1$ and $\log L_2$.
For the Gaussian copula, the BLF is obtained as
\begin{eqnarray}\label{eq:lf_gauss}
  \phi^{(2)}(L_1,L_2 ; \rho) \hspace{-3mm}&=&\hspace{-3mm} 
    \frac{1}{\sqrt{\det \vmatrix}}
    \exp\left\{-\frac{1}{2} \left[ \vpsi^T \left(\vmatrix^{-1} - \imatrix \right) \vpsi \right] \right\} 
    \phi^{(1)}_1 (L_1) \phi^{(1)}_2 (L_2) \;,
\end{eqnarray}
where 
\begin{eqnarray}
  \vpsi = \left[ \Psi^{-1} \left( \Phi^{(1)}_1 (L_1) \right) , \; \Psi^{-1} \left( \Phi^{(1)}_2 (L_2) \right) \right]^T
\end{eqnarray} 
and $\vmatrix$ is again defined by Equation~(\ref{eq:vmatrix}). 

\subsection{The FIR-FUV BLF}
\label{subsec:fir_fuv_blf}

Here, to make our model BLF astrophysically realistic, we construct
the FUV--FIR BLF by the copula method.
For the IR, we use the analytic form for the LF proposed by 
\citet{saunders90} which is defined as
\begin{eqnarray}\label{eq:saunders}
  \phi^{(1)}_1 (L) = \phi_{*1} \left( \frac{L}{L_{*1}} \right)^{1-\alpha_1}
    \exp \left\{ -\frac{1}{2\sigma^2} 
    \left[ \log \left(1+\frac{L}{L_{*1}}\right)\right]^2\right\}\;.
\end{eqnarray}
We adopt the parameters estimated by \citet{takeuchi03b} which are
obtained from the {\sl IRAS} PSC$z$ galaxies \citep{saunders00}.
For the UV, we adopt the Schechter function \citep{schechter76}.
\begin{eqnarray}\label{eq:schechter}
  \phi^{(1)}_2 (L) = (\ln 10)\; \phi_{*2} \left( \frac{L}{L_{*2}} \right)^{1-\alpha_2}
  \exp \left[-\left(\frac{L}{L_{*2}}\right)\right]\;,
\end{eqnarray}
We use the parameters presented by \citet{wyder05} for {\it GALEX} FUV
($\lambda_{\rm eff} = 1530\;$\AA): 
$(\alpha_2, L_{*2}, \phi_{*2}) = (1.21, 1.81\times 10^9h^{-2}\;L_\odot, 
1.35 \times 10^{-2}h^3\;\mbox{Mpc}^{-3})$.
For simplicity, we neglect the $K$-correction.
{ 
We use the re-normalized version of Eqs.~(\ref{eq:saunders}) and (\ref{eq:schechter}) 
so that they can be regarded as PDFs, as mentioned above.
}

\subsection{Result}
\label{subsec:result}

We show the constructed BLFs from the FGM and Gaussian copulas in Figures~\ref{fig:fgm_lf} 
and \ref{fig:gauss_lf}, respectively.
The FGM-based BLF cannot have a linear correlation coefficient larger than $\simeq 0.3$
as explained above, while the Gaussian-based BLF may have a much higher linear correlation.
We note that both copulas allow negative correlations, which are not discussed in this article.

First, even if the linear correlation coefficients are the same, the detailed structures of 
the BLFs with the FGM and Gaussian copulas are different 
(see the case of $\rho = 0.0 \mbox{--} 0.3$).
 For the Gaussian-based BLFs, we see a decline at the faint end, while
we do not have such structure in the FGM-based BLFs (see the closed contours in 
Fig.~\ref{fig:gauss_lf}).
This structure is introduced by the Gaussian copula, and from the physical point of view, 
it might not be strongly desired.
The FGM-based BLF has a more ideal shape.

Second, since the univariate LF shapes are different at FIR and FUV, the ridge of the
BLF is not a straight line but clearly nonlinear.
This feature is more clearly visible in higher correlation cases in Figure~\ref{fig:gauss_lf},
but always exists for the whole range of $\rho$.
This trend is indeed found in the $\lir$--$\luv$ diagram \citep{martin05}.
The underlying physics is that galaxies with high SFRs are more extinguished by
dust \citep[e.g.][]{buat07a,buat07b}.

Observational applications including this topic will be presented elsewhere (Takeuchi et al.\ 
2010, in preparation).

\section{Discussion}
\label{sec:discussion}

\subsection{Flux selection effect in multiband surveys}

\begin{figure*}
\centering\includegraphics[width=0.3\textwidth]{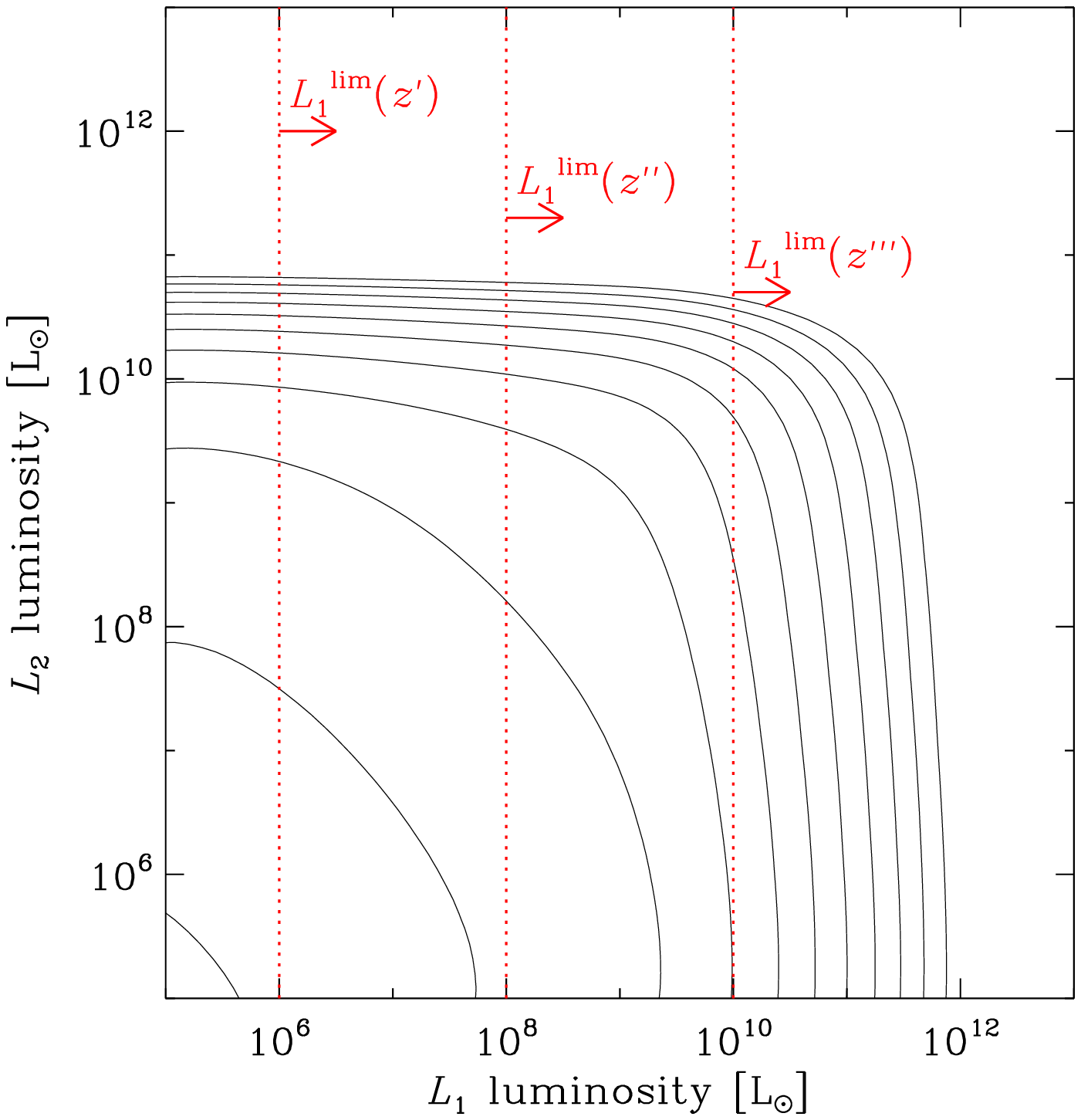}
\centering\includegraphics[width=0.3\textwidth]{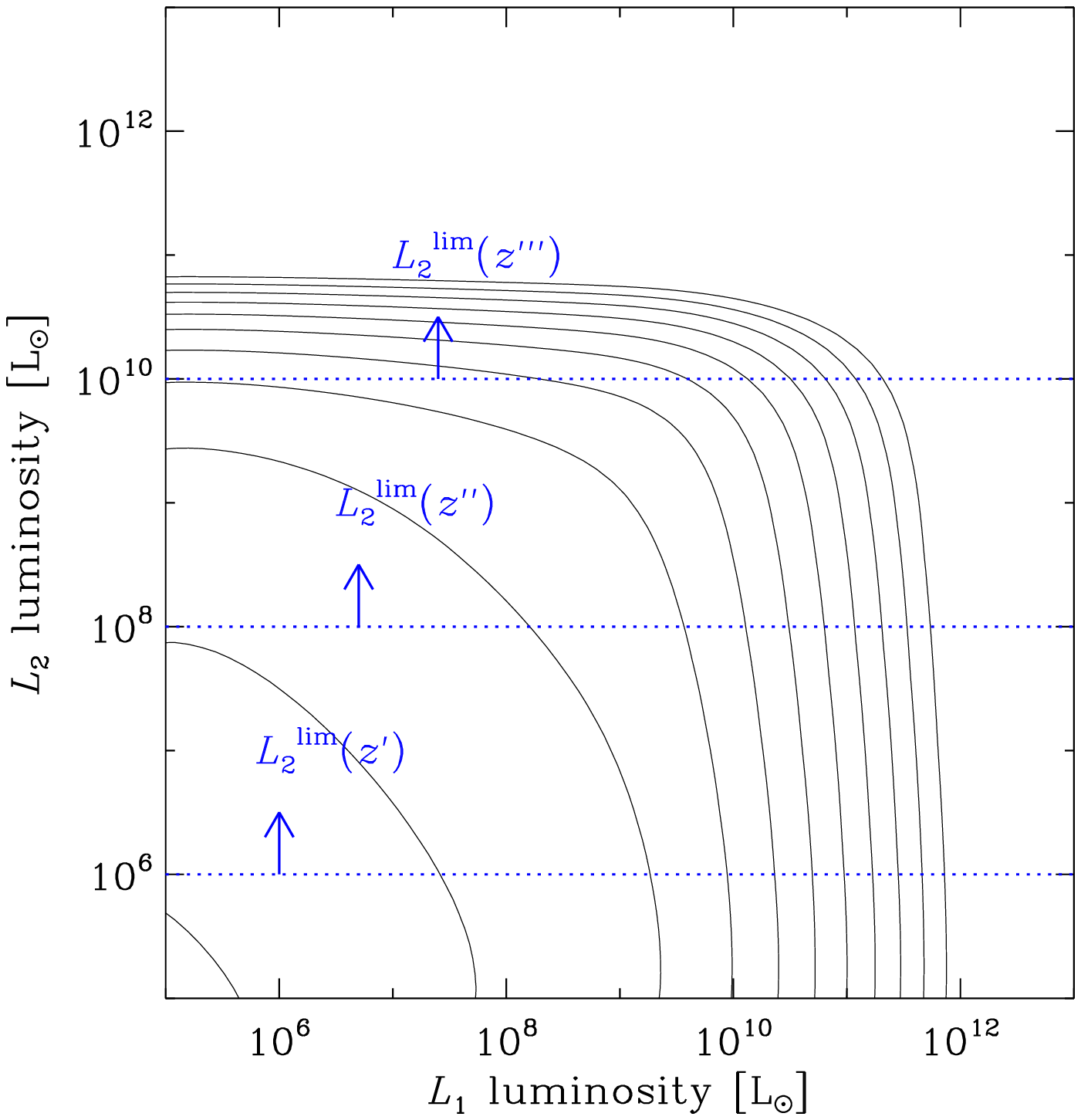}
\centering\includegraphics[width=0.3\textwidth]{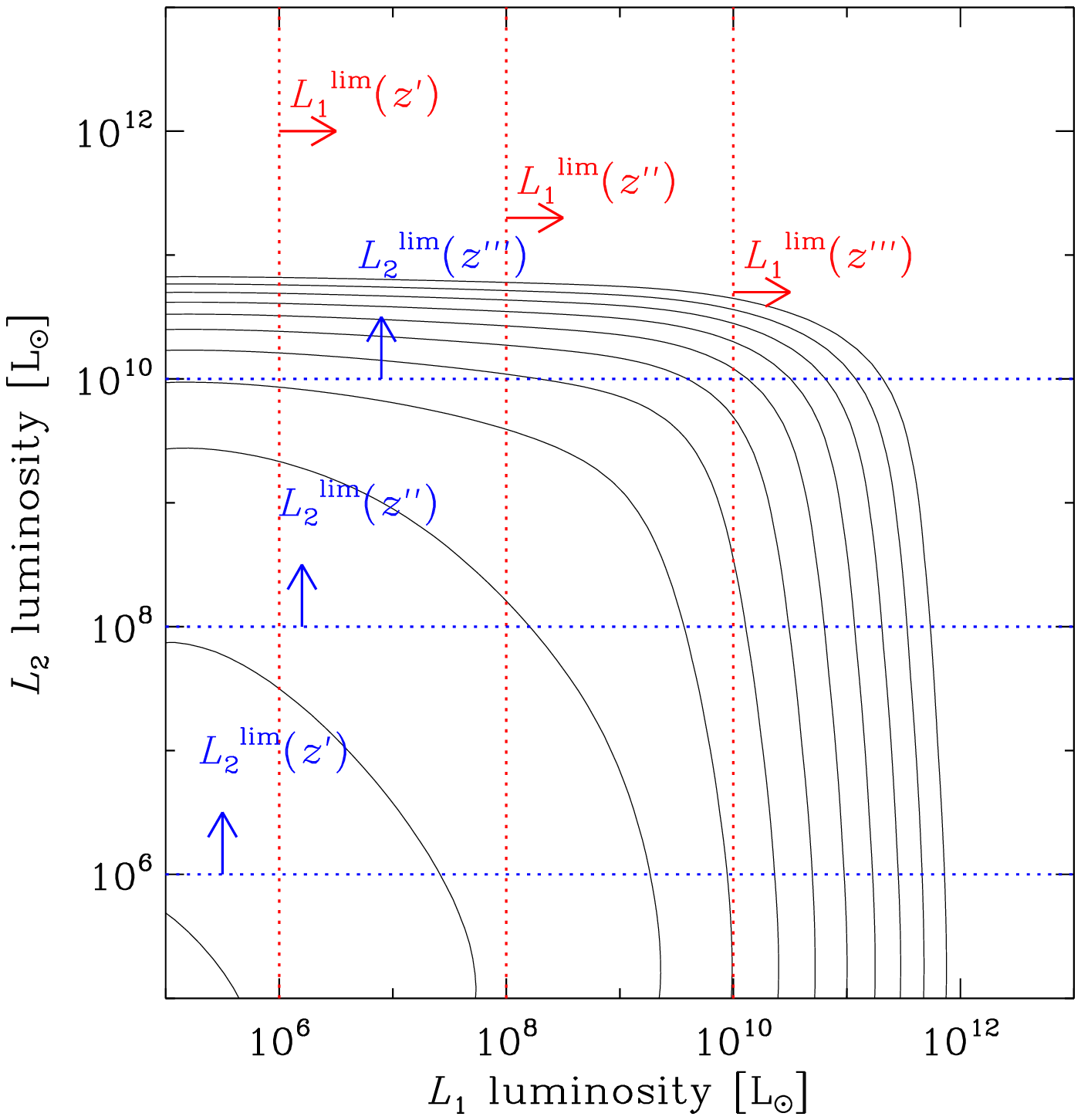}
\caption{Schematic description of the selection effect in a multiband survey.
Left: selection at band 1, center: selection at band 2, and
right: bivariate selection at band 1 and 2.}
\label{fig:selection_effect}
\end{figure*}

Since we have an explicit form of a BLF, we can discuss the flux selection effect
formally.
For simplicity, we consider the bivariate case (i.e.\ sample selected at two bands),
but it will be straightforward to extend the formulation to a multiwavelength case 
 (or more generally, selected using any physical properties).
The flux selection is described in terms of luminosity as putting a lower bound
$\llim$ on a luminosity--luminosity ($L_1$--$L_2$) plane.
The lower bound luminosity $\llim$ is defined by the flux (density) detection limit
$S^{\rm lim}$ as a function of redshift.
In most surveys, a certain wavelength band is chosen as the primary selection band,
like {\it B}-band, {\it K}s-band, $60\;\mu$m-selected, etc.
The schematic description of a survey is presented in Figure~\ref{fig:selection_effect}.

If we select a sample of objects (in our case galaxies) at band 1, the objects with
$L_1 < \llim_1 (z)$ would not be included in the sample at a certain redshift $z$.
Then, the detected sources should have $L_1 > \llim_1 (z)$ and $L_2 > \llim_2 (z)$.
Hence, on the $L_1$--$L_2$ plane, the 2-dim distribution of the detected objects 
is expressed as
\begin{eqnarray}\label{eq:detect}
  \sdet (L_1, L_2) \equiv \int_0^z \vel \phi^{(2)} \left( L_1 , L_2 \right) \Theta \left( \llim_1 (z') \right)   
  \Theta \left( \llim_2 (z') \right)  \pd z' \; ,
\end{eqnarray}
where $\Omega$ is a solid angle, and $\Theta$ is the Heaviside step function defined as
\begin{eqnarray}
  \Theta (a) = 
    \begin{cases}
    0 & L < a \\
    1 & L \geq a 
    \end{cases}
    \; .
\end{eqnarray}
The quantity $\sdet$ is proportional to the surface number density of objects detected in both 
bands on the $L_1$--$L_2$ plane.
We start from a primary selection at band 1, then we would have objects detected at band 1 
but not detected at band 2.
In such a case we only have upper limits for these objects.
The 2-dim distribution of the upper limits at band 2 is similarly formulated as
\begin{eqnarray}\label{eq:ul2}
  \sult (L_1, L_2) \equiv \int_0^z \vel \phi^{(2)} \left( L_1 , L_2 \right) 
    \Theta \left( \llim_1 (z') \right)
    \left[ 1 - \Theta \left( \llim_2 (z') \right)  \right]  \pd z' \; .
\end{eqnarray}
The superscript UL2 stands for ``upper limit at band~2''.
In statistical terminology, the upper limit case, i.e.\ we know there is an object but we do
only have the upper (or lower) limits of a certain quantity, is referred to as ``censored''.
Though we can define the distribution $\sult (L_1, L_2)$ by Eq.~(\ref{eq:ul2}), since the 
sample objects belonging to this category appear only as upper limits on the plot, 
a special statistical treatment, referred to as the survival analysis, is required to estimate 
$\sult (L_1, L_2)$ from the data.  
Since we select objects at band~1, we do not have upper limits at band~1, because 
we do not know if there would be an object below the limit.
This case is called ``truncated'' in statistics.

If we select objects at band~2, we can formulate the 2-dim distribution of detected objects
and upper limits exactly in the same way as the band~1 selected sample.
For the objects detected at both bands, the 2-dim distribution is expressed by Eq.~(\ref{eq:detect}).
The objects detected at band~2 but not detected at band~1 is expressed as 
\begin{eqnarray}\label{eq:ul1}
  \sulo (L_1, L_2) \equiv \int_0^z \vel \phi^{(2)} \left( L_1 , L_2 \right) 
    \left[ 1 - \Theta \left( \llim_1 (z') \right)  \right]  
    \Theta \left( \llim_2 (z') \right) \pd z' \; .
\end{eqnarray}

If we can model $\llim_1 (z)$ and $\llim_2 (z)$ precisely including the $K$-correction, evolutionary
effect, etc., we can use the observed bivariate luminosity distribution to estimate the
correlation coefficient, or more generally the dependence structure of two luminosities 
through Eqs.~(\ref{eq:detect})--(\ref{eq:ul2}).
We can deal with these cases in a unified manner with techniques developed in survival
analysis.
We discuss this issue in a subsequent work (Takeuchi et al.\ 2010, in preparation).

\subsection{Other possible applications}
\label{subsec:other}

\subsubsection{The star formation rate function}

The star formation rate (SFR) is one of the most fundamental quantities to investigate
the formation and evolution of galaxies.
The SFR is often estimated from the FUV flux of galaxies (or other related
observables like H$\alpha$ etc.) after ``correcting'' the dust extinction. 
However, some problems have been pointed out for this method.
For instance, the relation between the UV slope $\beta$ (or equivalently, FUV$-$NUV color)
and the FIR-FUV flux ratio $\lir/\luv$ (often referred to as the IRX-$\beta$ relation) 
is frequently used to correct the extinction,
but this relation is not always the same for various categories of star-forming galaxies
\citep[e.g.][]{buat05,boissier07,boquien09,takeuchi10a}.

Instead, the total SFR obtained from the FUV and FIR luminosities would be a more
reliable measure of the SFR since both are directly observable values
\citep[e.g.][]{iglesias04, buat05, iglesias06, buat07a, buat07b, takeuchi10a}. 
Assuming a constant SFR over $10^8 \mbox{yr}$, and Salpeter initial mass function 
\citep[][mass range: $0.1\mbox{--}100\;M_\odot$]{salpeter55}, 
we have the relation between the SFR and $\luv$ 
\begin{eqnarray}
  \log \sfruv = \log \luv  - 9.51 \;.
\end{eqnarray}
For the FIR, to transform the dust emission to the SFR, we assume
that all the stellar light is absorbed by dust. 
Then, we obtain the following formula under the same assumption for
both the SFR history and the IMF as those of the FUV,
\begin{eqnarray}
  \log \sfrir = \log \ltir - 9.75 - \log (1-\eta) \;.
\end{eqnarray}
Here, $\eta$ is the fraction of the dust emission by old stars which is not
related to the current SFR \citep{hirashita03}, and $\ltir$ is the FIR luminosity
integrated over $\lambda = 8 \mbox{--} 1000\;\mu$m.
Thus, the total SFR is simply
\begin{eqnarray}
  \sfr = \sfruv + \sfrir
\end{eqnarray}
\citep{iglesias06}.
Since the total SFR is basically estimated from the luminosities at FUV and FIR
(note that $\sfruv \propto \luv$ and $\sfrir \propto \ltir$), 
the estimation of the PDF of the total SFR reduces to the estimation of the FIR-FUV
BLF \citep[e.g.][]{takeuchi10b}.
{ 
However, since the total SFR is the sum of two {\it dependent} variables, it is not
straightforward to formulate the function unlike the cases we have seen above.
This analysis will be discussed with a more specific methodology in our future work.
}

\subsubsection{The distribution of the specific star formation rate}

Another direct application is the distribution of the specific SFR (SSFR), $\mbox{SFR}/M_*$
where $M_*$ is the total stellar mass of a galaxy.
The SSFR has gained much attention in the last decade, since the relation between 
$M_*$ and the SSFR of galaxies turns out to be a very important clue to understand 
the SF history of galaxies: more massive galaxies have ceased their SF activity earlier 
in the cosmic time than less massive galaxies 
\citep[downsizing in redshift: e.g.][among others]{cowie96,boselli01,heavens04,
feulner05,noeske07a,noeske07b,panter07,damen09a,damen09b}.
For a comprehensive summary of the downsizing, readers are encouraged to read
Introduction of \citet{fontanot09}.
Despite of its importance, the treatment of multiwavelength data for this analysis is
inevitably complicated and does not seem to be well understood, because we must deal 
with the data related to SFR and $M_*$ estimation.
This might be, at least partially, the reason why the quantitative values of the $M_*$--SSFR relation
are different among different studies.

As may easily guess after the above discussions, the $M_*$--SSFR relation can be
reduced to the relation between a luminosity at a certain mass-related band (often near IR bands) 
and a SF-related one (FUV, FIR, etc.)
Then, we can model, for example, a $L_K$--$\ltir$ bivariate luminosity function ($L_K$: 
{\it K}-band luminosity)\footnote{More precisely, $L_K$--$\ltir$--$\luv$ trivariate function 
might be appropriate for this issue.} to examine the observed relation including 
all the selection effects.
This is particularly useful for this topic, since \citet{takeuchi10b} found that the SFRF 
cannot be described by the Schechter function unlike the assumptions adopted in previous 
studies, but much more similar to the Saunders IR LF [Eq.~(\ref{eq:saunders})].
The selection effect would be more complicated than in the case of the same Schechter
marginals, but can be treated in the same way as discussed above. 
Thus, this may also be an interesting application of the copula-based BLF in the
epoch of future large surveys.

\section{Summary and Conclusions}\label{sec:conclusion}

In this work, we introduced an analytic method to construct a bivariate distribution 
function (DF) with given marginal distributions and correlation coefficient,
by making use of a convenient mathematical tool, called a copula.
Using this mathematical tool, we presented an application to construct a bivariate 
LF of galaxies (BLF).
Specifically, we focused on the FUV--FIR BLF, since these two luminosities are 
related to the star formation (SF) activity.
Though both the FUV and FIR are related to the SF activity, the marginal univariate 
LFs have a very different functional form: former is well described by Schechter function 
whilst the latter has a much more extended power-law like luminous end.
We constructed the FUV-FIR BLFs by the FGM and Gaussian copulas with different
strength of correlation, and examined their statistical properties.
Then, we discussed some further possible applications of the BLF: the problem of 
a multiband flux-limited sample selection, the construction of the SF rate (SFR)
function, and the construction of the stellar mass of galaxies ($M_*$)--specific SFR
($\mbox{SFR}/M_*$) relation.

We summarize our conclusions as follows:
\begin{enumerate}
\item If the correlation of two variables is weak (Pearson's correlation coefficient 
$|\rho| <1/3 $), the Farlie-Gumbel-Morgenstern (FGM) copula
provides an intuitive and natural way for constructing such a bivariate DF.
\item When the linear correlation is stronger, the FGM copula becomes inadequate, 
in which case a Gaussian copula should be preferred. 
The latter connects two marginals and is directly related to the linear correlation 
coefficient between two variables.
\item Even if the linear correlation coefficient is the same, the structure of a
BLF is different depending on the choice of a copula. 
Hence, a proper copula should be chosen for each case.
\item The model FIR-FUV BLF was constructed. 
Since the functional shape of the LF at each wavelength is very different, 
the obtained BLF has a clear nonlinear structure.
This feature was indeed found in actual observational data \citep[e.g.,][]{martin05}.
\item We formulated the problem of the multiwavelength selection effect by the BLF.
This enables us to deal with datasets derived from surveys presenting complex 
selection functions.
\item We discussed the estimation of the SFR function (SFRF) of galaxies.
The copula-based BLF will be a convenient tool to extract detailed information from
the observationally estimated SFRF because of its bivariate nature.
\item The stellar mass--specific SFR relation was also discussed.
This relation can be reduced to a BLF of luminosities at a mass-related band and
a SF-related band.
With an analytic BLF model constructed by a copula will provide us with a 
powerful tool to analyze the downsizing phenomenon with addressing the
complicated selection effects.
\end{enumerate}

As the copula becomes better known to the astrophysical community and statisticians 
develop the copula functions, we envision many more interesting applications in
the future. 
In a series of forthcoming papers, we will present more observationally-oriented 
applications of copulas.

\section*{Acknowledgements}
{ 
First we thank the anonymous referee for her/his careful reading of the manuscripts
and many suggestions which improved this paper.
}
We are grateful to V\'{e}ronique Buat, Bruno Milliard, Agnieszka Pollo, Akio K.\ Inoue, 
Denis Burgarella, Kiyotomo Ichiki, and Masanori Sato for enlightening discussions.
We also thank Masanori Sato and Takako T.\ Ishii for helping the development of 
the numerical routines to calculate copulas.
We have been supported by Program for Improvement of Research 
Environment for Young Researchers from Special Coordination Funds for 
Promoting Science and Technology, and the Grant-in-Aid for the Scientific 
Research Fund (20740105) commissioned by the Ministry of Education, Culture, 
Sports, Science and Technology (MEXT) of Japan.
We are also partially supported from the Grand-in-Aid for the Global 
COE Program ``Quest for Fundamental Principles in the Universe: from 
Particles to the Solar System and the Cosmos'' from the MEXT.

\appendix

\section{An extension of the FGM system by Johnson \& Kotz (1977)}
\label{sec:jk77}

\begin{figure*}
\centering\includegraphics[angle=0,width=0.3\textwidth]{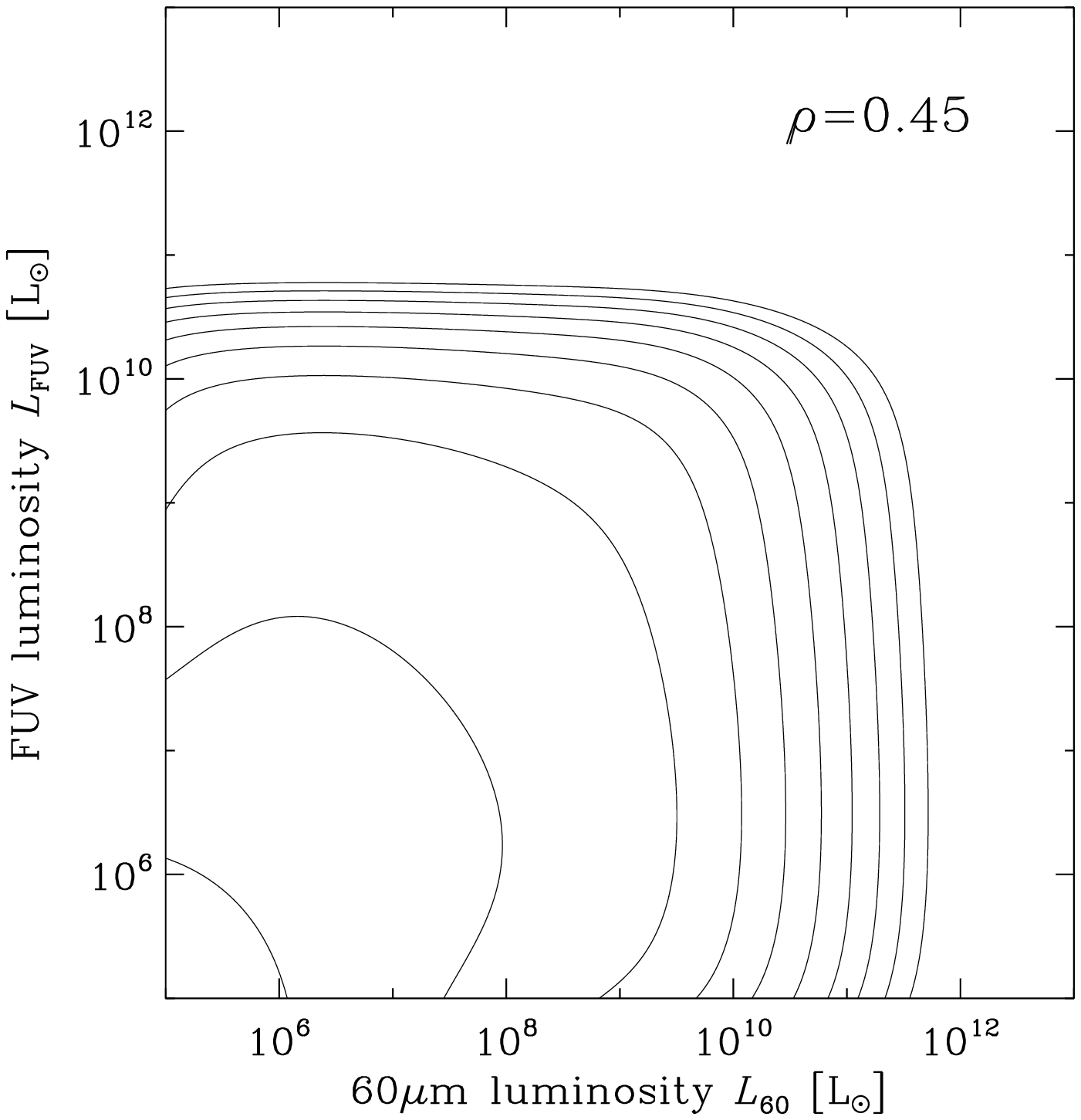}
\caption{The BLF constructed with a first-order iterated FGM copula.
The linear correlation coefficient $\rho$ is 0.45.}
\label{fig:fgm_lf_iter}
\end{figure*}

Although the FGM system of distributions provides us with convenient tool to 
construct the statistical model, its usefulness is restricted
by the limitation of the correlation strength described above.
{}To overcome this drawback, many attempts have been made to extend the
FGM distributions \citep[see, e.g.,][]{stuart94,kotz00}.
Among them, \citet{johnson77} introduced the following iterated generalization
of Equation~(\ref{eq:cum_fgm}):
\begin{eqnarray}\label{eq:cum_jk}
  G(x_1,x_2) =
    \sum_{j=0}^{k} \kappa_j \left[F_1(x_1)F_2(x_2)\right]^{[j/2]+1} 
    \left\{\left[1-F_1(x_1)\right]\left[1-F_2(x_2)\right]\right\}^{[(j+1)/2]}
    \;,
\end{eqnarray}
where the symbol in the exponent $[j/2]$ means the maximum natural number which
does not exceed $j/2$.
We set $\kappa_0=1$.
\citet{huang84} examined the dependence structure of 
Equation~(\ref{eq:cum_jk}) especially for the case of $k=2$, 
and showed that the correlation can be stronger for these extension.
In the case of the one-iteration family ($k=2$), we have the DF as
\begin{eqnarray}\label{eq:cum_hk}
  G(x_1,x_2) = F_1(x_1)F_2(x_2) \left\{ 1
    + \kappa_1 \left[1-F_1(x_1)\right]\left[1-F_2(x_2)\right]
    + \kappa_2 F_1(x_1)F_2(x_2)
      \left[1-F_1(x_1)\right]\left[1-F_2(x_2)\right]\right\} \;.
\end{eqnarray}
The corresponding PDF is
\begin{eqnarray}\label{eq:dif_hk}
  g(x_1,x_2) =f_1(x_1)f_2(x_2) \left\{ 1
    + \kappa_1 \left[2F_1(x_1)-1\right]\left[2F_2(x_2)-1\right]
    + \kappa_2 F_1(x_1)F_2(x_2)
      \left[3F_1(x_1)-2\right]\left[3F_2(x_2)-2\right]\right\} \;.
\end{eqnarray}
Then, just the same as the case of the original FGM distribution ($k=1$), 
we obtain the covariance 
\begin{eqnarray}\label{eq:cov_hk}
  \mathsf{Cov}(x_1,x_2)
  \hspace{-3mm}&=&\hspace{-3mm} \iint (x_1 - \bar{x}_1)(x_2 - \bar{x}_2)
    f_1(x_1)f_2(x_2) \nonumber \\
    &&\times \left\{ 1
    + \kappa_1 \left[2F_1(x_1)-1\right]\left[2F_2(x_2)-1\right] 
    + \kappa_2 F_1(x_1)F_2(x_2)
      \left[3F_1(x_1)-2\right]\left[3F_2(x_2)-2\right]\right\} dx_1 dx_2
    \nonumber \\
  \hspace{-3mm}&=&\hspace{-3mm} 
    \kappa_1 \int x_1 f_1(x_1)\left[2F_1(x_1)-1\right] dx_1
      \int x_2 f_2(x_2) \left[2F_2(x_2)-1\right] dx_2 \nonumber \\
  &&+ \kappa_2 \int x_1 f_1(x_1) F_1(x_1) \left[3F_1(x_1)-2\right] dx_1
      \int x_2 f_2(x_2) F_2(x_2) \left[3F_2(x_2)-2\right] dx_2
    \nonumber \\
  \hspace{-3mm}&\equiv&\hspace{-3mm} 
    A_1 \kappa_1 + A_2 \kappa_2  \;.
\end{eqnarray}
\citet{huang84} obtained the parameter space for $\kappa_1$ and $\kappa_2$ 
as\footnote{Note that there is a typo for Equation~(\ref{eq:param_hk2})
in \citet{huang84}.
There are also typos in Equations~(\ref{eq:param_hk2}) and (\ref{eq:param_hk3})
in \citet{kotz00}.
}
\begin{eqnarray}
  &&\hspace{-6mm} \left| \kappa_1 \right| \le 1 \label{eq:param_hk1}\;,\\
  &&\hspace{-6mm} \kappa_1 + \kappa_2 \ge -1 \label{eq:param_hk2}\;, \\
  &&\hspace{-6mm} \kappa_2 \le \frac{3 - \kappa_1 + 
    \sqrt{3 (1-\kappa_1)(3+\kappa_1)}}{2}\;.\label{eq:param_hk3}
\end{eqnarray}
They showed that, for a positive correlation, 
\begin{eqnarray}
  \rho \le \frac{\kappa_1}{3} + \frac{31 \kappa_2}{240}  \;.
\end{eqnarray}
Under these conditions, we have $\rho \le 0.5072$, which is considerably 
better than $1/3$.
The BLF constructed with the first-order iterated FGM copula is expressed as
\begin{eqnarray}\label{eq:lf_hk}
  \phi^{(2)}(L_1,L_2) \hspace{-3mm}&=&\hspace{-3mm} \phi^{(1)}_1(L_1)\phi^{(1)}_2(L_2) \nonumber \\
    &&\times \left\{ 1
    + \kappa_1 \left[2\Phi^{(1)}_1 (L_1)-1\right]\left[2\Phi^{(1)}_2 (L_2)-1\right]
    + \kappa_2 \phi^{(1)}_1(L_1)\phi^{(1)}_2(L_2)
      \left[3\Phi^{(1)}_1(L_1)-2\right]\left[3\Phi^{(1)}_2(L_2)-2\right]\right\} \;.
\end{eqnarray}
The FIR-FUV BLF by Equation~{\ref{eq:lf_hk}} is shown in Figure~\ref{fig:fgm_lf_iter}.
Clearly the dependence between the two luminosities is stronger than the original
FGM-based BLF.
However, now it is not intuitive nor straightforward to relate these two parameters of 
dependence $\kappa_1$ and $\kappa_2$ to the linear correlation coefficient. 

\section{Estimators of the nonparametric dependence measures $\rs$ and $\tau$}
\label{sec:nonparam_est}

Here we present the estimators of the nonparametric measure of dependence introduced in
Section~\ref{subsec:nonparam}. 
More detailed derivation and properties of these nonparametric measures of dependence are found in e.g., 
\citet{hettmansperger84} and \citet{hollander99}.

Let $\{R_i\}_{i=1,\dots,n}$ and $\{S_i\}_{i=1,\dots,n}$ be the ranks of $\{x_{1i}\}_{i=1,\dots,n}$ and 
$\{x_{2i}\}_{i=1,\dots,n}$, respectively.
If we denote the estimator of $\rs$ for a sample as $r_{\rm S}$, 
\begin{eqnarray}
  r_{\rm S} = \frac{\mathstrut \sum_{i=1}^n \left( R_i - \frac{n+1}{2} \right)\left( S_i - \frac{n+1}{2} \right)}{
    \mathstrut \sqrt{\sum_{i=1}^n \left( R_i - \frac{n+1}{2} \right) \sum_{i=1}^n\left( S_i - \frac{n+1}{2} \right)}} \;.
\end{eqnarray}
This is also expressed as
\begin{eqnarray}\label{eq:spearman_sample}
  r_{\rm S} = \frac{\mathstrut 12 \sum_{i=1}^n \left( R_i - \frac{n+1}{2} \right)\left( S_i - \frac{n+1}{2} \right)}{
    \mathstrut n(n^2-1)} \;.
\end{eqnarray}
This form is an exact sample counterpart of Equation~(\ref{eq:spearman_copula}).
If we define $d_i \equiv S_i - R_i$, Equation~(\ref{eq:spearman_sample}) reduces to the following simper 
form
\begin{eqnarray}
  r_{\rm S} = 1 - \frac{\mathstrut 6 \sum_{i=1}^n d_i^2}{\mathstrut n(n^2-1)} \;.
\end{eqnarray}
The variance of $r_{\rm S}$ in the large-sample limit is given by
\begin{eqnarray}
  \mathsf{Var} [r_{\rm S}] = \frac{1}{n-1} \;.
\end{eqnarray}

The most basic form of Kendall's $\tau$ for a sample has been already shown in 
Section~\ref{subsec:nonparam} [Equation~(\ref{eq:kendall_sample})].
It is also expressed as
\begin{eqnarray}
  t \hspace{-3mm}&=&\hspace{-3mm} \frac{\mathstrut 2(\nc - \nd)}{\mathstrut n(n-1)} \nonumber \\
   \hspace{-3mm}&=&\hspace{-3mm} 1-\frac{\mathstrut 4}{\mathstrut n(n-1)} \nd 
\end{eqnarray}  
since $\nc + \nd = n(n-1)/2$.
The variance of $t$ is given by 
\begin{eqnarray}
  \mathsf{Var}[t] = \frac{2(2n+5)}{9n(n-1)} 
\end{eqnarray}
\citep[see ][for a very concise derivation]{valz90}.


\begin{thebibliography}{} 
\bibitem[\protect\citeauthoryear{Ball et al.}{2006}]{ball06}
  Ball N.\ M., Loveday J., Brunner R.\ J., Baldry I.\ K., Brinkmann J.,  
  2006, MNRAS, 373, 845

\bibitem[\protect\citeauthoryear{Binggeli, Sandage, \& Tammann}{1988}]{binggeli88}
  Binggeli B., Sandage A., Tammann G.~A., 1988, ARA\&A, 26, 509 

\bibitem[\protect\citeauthoryear{Blanton et al.}{2001}]{blanton01}
  Blanton M.~R., et al., 2001, AJ, 121, 2358 

\bibitem[\protect\citeauthoryear{Benabed et al.}{2009}]{benabed09}
  Benabed K., Cardoso J.-F., Prunet S., Hivon E., 2009, MNRAS, 400, 219 

\bibitem[\protect\citeauthoryear{Boissier et al.}{2007}]{boissier07}
  Boissier S., et al., 2007, ApJS, 173, 524 

\bibitem[\protect\citeauthoryear{Boquien et al.}{2009}]{boquien09}
  Boquien M., et al., 2009, ApJ, 706, 553 

\bibitem[\protect\citeauthoryear{Boselli et al.}{2001}]{boselli01}
  Boselli A., Gavazzi G., Donas J., Scodeggio M., 2001, AJ, 121, 753 

\bibitem[\protect\citeauthoryear{Buat \& Burgarella}{1998}]{buat98}
  Buat V., Burgarella D., 1998, A\&A, 334, 772

\bibitem[\protect\citeauthoryear{Buat et al.}{2005}]{buat05} 
  Buat V., et al., 2005, ApJ, 619, L51
 
\bibitem[\protect\citeauthoryear{Buat et al.}{2007a}]{buat07a} 
  Buat V., et al., 2007a, ApJS, 173, 404 

\bibitem[\protect\citeauthoryear{Buat et al.}{2007b}]{buat07b}
  Buat V., Marcillac D., Burgarella D., Le Floc'h E., Takeuchi T.~T., Iglesias-P{\'a}ramo J., 
  Xu C.~K., 2007b, A\&A, 469, 19 

\bibitem[\protect\citeauthoryear{Buat et al.}{2008}]{buat08}
  Buat V., et al., 2008, A\&A, 483, 107 

\bibitem[\protect\citeauthoryear{Buat et al.}{2009}]{buat09}
  Buat V., Takeuchi T.~T., Burgarella D., Giovannoli E., Murata K.~L., 2009, A\&A, 507, 693 

\bibitem[\protect\citeauthoryear{Cambanis}{1977}]{cambanis77}
  Cambanis S., 1977, J.\ Multivariate Analysis, 7. 551

\bibitem[\protect\citeauthoryear{Chapman et al.}{2003}]{chapman03}
  Chapman S.~C., Helou G., Lewis G.~F., Dale D.~A., 2003, ApJ, 588, 186 

\bibitem[\protect\citeauthoryear{Cho{\l}oniewski}{1985}]{choloniewski85} 
  Cho{\l}oniewski J., 1985, MNRAS, 214, 197 

\bibitem[\protect\citeauthoryear{Cowie et al.}{1996}]{cowie96} 
  Cowie L.~L., Songaila A., Hu E.~M., Cohen J.~G., 1996, AJ, 112, 839 

\bibitem[\protect\citeauthoryear{Cross \& Driver}{2002}]{cross02}
  Cross N., Driver S.~P., 2002, MNRAS, 329, 579 

\bibitem[\protect\citeauthoryear{Damen et al.}{2009a}]{damen09a} 
  Damen M., Labb{\'e} I., Franx M., van Dokkum P.~G., Taylor E.~N., Gawiser 
  E.~J., 2009a, ApJ, 690, 937 

\bibitem[\protect\citeauthoryear{Damen et al.}{2009b}]{damen09b} 
  Damen M., F{\"o}rster Schreiber N.~M., Franx M., Labb{\'e} I., Toft S., van 
  Dokkum P.~G., Wuyts S., 2009b, ApJ, 705, 617 

\bibitem[\protect\citeauthoryear{de Lapparent et al.}{2003}]{delapparent03}
  de Lapparent V., Galaz G., Bardelli S., Arnouts S., 2003, A\&A, 404, 831 

\bibitem[\protect\citeauthoryear{D'Este}{1981}]{deste81}
  D'Este G.\ M., 1981, Biometrika, 68, 339
  
\bibitem[\protect\citeauthoryear{Driver et al.}{2006}]{driver06}
  Driver S.\ P., et al., 2006, MNRAS, 368, 414

\bibitem[\protect\citeauthoryear{Farlie}{1960}]{farlie60}
  Farlie D.\ J.\ G., 1960, Biometrika, 47, 307

\bibitem[\protect\citeauthoryear{Feulner et al.}{2005}]{feulner05}
  Feulner G., Gabasch A., Salvato M., Drory N., Hopp U., Bender R., 2005, ApJ, 633, L9 

\bibitem[\protect\citeauthoryear{Fontanot et al.}{2009}]{fontanot09}
  Fontanot F., De Lucia G., Monaco P., Somerville R.~S., Santini P., 2009, MNRAS, 397, 1776 

\bibitem[\protect\citeauthoryear{Gumbel}{1960}]{gumbel60}
  Gumbel E.\ J., 1960, J.\ Amer.\ Statist.\ Assoc., 55, 698

\bibitem[\protect\citeauthoryear{Heavens et al.}{2004}]{heavens04}
  Heavens A., Panter B., Jimenez R., Dunlop J., 2004, Natur, 428, 625 

\bibitem[\protect\citeauthoryear{Hettmansperger}{1984}]{hettmansperger84}
  Hettmansperger T.\ P., 1984, Statistical Inference Based on Ranks, John Wiley \& Sons, New York

\bibitem[\protect\citeauthoryear{Hirashita, Buat, \& Inoue}{2003}]{hirashita03}
  Hirashita H., Buat V., Inoue A.~K., 2003, A\&A, 410, 83 

\bibitem[\protect\citeauthoryear{Hollander \& Wolfe}{1999}]{hollander99}
  Hollander M., \& Wolfe D.\ A., 1999, Nonparametric Statistical Methods, 2nd ed., John Wiley \& 
  Sons, New York

\bibitem[\protect\citeauthoryear{Huang \& Kotz}{1984}]{huang84}
  Huang J.\ S., Kotz S., 1984, Biometrika, 71, 633

\bibitem[\protect\citeauthoryear{Iglesias-P{\'a}ramo et al.}{2004}]{iglesias04}
  Iglesias-P{\'a}ramo J., Buat V., Donas J., Boselli A., Milliard B., 2004, A\&A, 419, 109 
  
\bibitem[\protect\citeauthoryear{Iglesias-P{\'a}ramo et al.}{2006}]{iglesias06}
  Iglesias-P{\'a}ramo J., et al., 2006, ApJS, 164, 38 

\bibitem[\protect\citeauthoryear{Jiang et al.}{2009}]{jiang09} 
  Jiang I.-G., Yeh L.-C., Chang Y.-C., Hung W.-L., 2009, AJ, 137, 329 

\bibitem[\protect\citeauthoryear{Johnson \& Kotz}{1977}]{johnson77}
  Johnson N.\ L., Kotz S., 1977, Comm.\ Statist.\ Ser.\ A (Theory and Methods),
  6, 485

\bibitem[\protect\citeauthoryear{Koen}{2009}]{koen09}
  Koen C., 2009, MNRAS, 393, 1370 

\bibitem[\protect\citeauthoryear{Kotz, Balakrishnan, \& Johnson}{2000}]{kotz00}
  Kotz S., Balakrishnan N., Johnson N., L., 2000, Continuous Multivariate
  Distributions, Volume 1: Models and Applications, 2nd ed., John Wiley \& Sons,
  New York, pp.51--62

\bibitem[\protect\citeauthoryear{Lin et al.}{1996}]{lin96}
  Lin H., Kirshner R.~P., Shectman S.~A., Landy S.~D., Oemler A., Tucker 
  D.~L., Schechter P.~L., 1996, ApJ, 464, 60 

\bibitem[\protect\citeauthoryear{Martin et al.}{2005}]{martin05} 
  Martin D.~C., et al., 2005, ApJ, 619, L59 

\bibitem[\protect\citeauthoryear{Mobasher, Sharples, \& Ellis}{Mobasher et al.}{1993}]{mobasher93}
  Mobasher B., Sharples R.\ M., Ellis R.\ S., 1993, MNRAS, 263, 560

\bibitem[\protect\citeauthoryear{Morgenstern}{1956}]{morgenstern56}
  Morgenstern D., 1956, Mitt.\ Math.\ Stat., 8, 234

\bibitem[\protect\citeauthoryear{Nelsen}{2006}]{nelsen06}
  Nelsen R. B., 2006, An Introduction to Copulas, 2nd ed., Springer, New York, 
  \S 2

\bibitem[\protect\citeauthoryear{Noeske et al.}{2007a}]{noeske07a} 
  Noeske K.~G., et al., 2007a, ApJ, 660, L43 

\bibitem[\protect\citeauthoryear{Noeske et al.}{2007b}]{noeske07b} 
  Noeske K.~G., et al., 2007b, ApJ, 660, L47 

\bibitem[\protect\citeauthoryear{Panter et al.}{2007}]{panter07} 
  Panter B., Jimenez R., Heavens A.~F., Charlot S., 2007, MNRAS, 378, 1550 

\bibitem[\protect\citeauthoryear{Salpeter}{1955}]{salpeter55}  
  Salpeter E.~E., 1955, ApJ, 121, 161 

\bibitem[\protect\citeauthoryear{Saunders et al.}{1990}]{saunders90}
  Saunders W., Rowan-Robinson M., Lawrence A., Efstathiou G., Kaiser N., 
  Ellis R.~S., Frenk C.~S., 1990, MNRAS, 242, 318 
 
\bibitem[\protect\citeauthoryear{Saunders et al.}{2000}]{saunders00}
  Saunders W., et al., 2000, MNRAS, 317, 55 
 
\bibitem[\protect\citeauthoryear{Schechter}{1976}]{schechter76}
  Schechter P.\ L., 1976, ApJ, 203, 297

\bibitem[\protect\citeauthoryear{Schucany, Parr, \& Boyer}{1978}]{schucany78}
  Schucany W.\ R., Parr W.\ C., Boyer J.\ E., 1978, Biometrika, 65, 650

\bibitem[\protect\citeauthoryear{Schafer}{2007}]{schafer07}
  Schafer C.\ M., 2007, ApJ, 661, 703

\bibitem[\protect\citeauthoryear{Scherrer et al.}{2010}]{scherrer10}
  Scherrer R.~J., Berlind A.~A., Mao Q., McBride C.~K., 2010, ApJ, 708, L9 

\bibitem[\protect\citeauthoryear{Sklar}{1959}]{sklar59}
  Sklar A., 1959, Publ. Inst. Stat. Univ. Paris, 8, 229
  
\bibitem[\protect\citeauthoryear{Stuart \& Ord}
{Stuart et al.}{1994}]{stuart94}
  Stuart A., Ord K., 1994, Kendall's Advanced Theory of Statistics, 6th
  ed.\ Vol.\ 1, Distribution Theory, Arnold, London, pp.275--276

\bibitem[\protect\citeauthoryear{Takeuchi}{2000}]{takeuchi00a}
  Takeuchi T.\ T., 2000, Ap\&SS, 271, 213

\bibitem[\protect\citeauthoryear{Takeuchi, Yoshikawa, \& Ishii}
{Takeuchi et al.}{2000}]{takeuchi00b}
  Takeuchi T.\ T., Yoshikawa, K., Ishii, T.\ T., 2000, ApJS, 129, 1

\bibitem[\protect\citeauthoryear{Takeuchi, Yoshikawa, \& Ishii}
{Takeuchi et al.}{2003b}]{takeuchi03b}
  Takeuchi T.\ T.,  Yoshikawa K., Ishii T.\ T., 2003, ApJ, 587, L89
 
\bibitem[\protect\citeauthoryear{Takeuchi, Buat, \& Burgarella}{Takeuchi et al.}{2005c}]{takeuchi05c}
  Takeuchi T.~T., Buat V., Burgarella D., 2005, A\&A, 440, L17 
 
\bibitem[\protect\citeauthoryear{Takeuchi et al.}{2010a}]{takeuchi10a}
  Takeuchi T.~T., Buat V., Heinis S., Giovannoli E., Yuan F.~-T., 
  Iglesias-Paramo J., Murata K.~L., Burgarella D., 2010a, A\&A, in press
  (astro-ph/0912.5051)
 
\bibitem[\protect\citeauthoryear{Takeuchi et al.}{2010b}]{takeuchi10b}
  Takeuchi T.~T., Buat V., Burgarella E., Giovannoli E., Murata K.~L., 
  Iglesias-P\'{a}ramo J., Hern\'{a}ndez-Fern\'{a}ndez J., 2010b, in Hunting for the Dark: the Hidden
  Side of Galaxy Formation, AIP Conf.\ Ser., in press
 
\bibitem[\protect\citeauthoryear{Trivedi \& Zimmer}{2005}]{trivedi05}
  Trivedi P.\ R., Zimmer D.\ M., 2005, Foundations and Trends in Econometrics, 1, 1
 
\bibitem[\protect\citeauthoryear{Valz \& Mcleod}{1990}]{valz90}
  Valz P.\ D., \& McLeod A.\ I., 1990, Amer.\ Statist., 44, 39

\bibitem[\protect\citeauthoryear{Willmer et al.}{2006}]{willmer06}
  Willmer C.~N.~A., et al., 2006, ApJ, 647, 853 

\bibitem[\protect\citeauthoryear{Wyder et al.}{2005}]{wyder05} 
  Wyder T.~K., et al., 2005, ApJ, 619, L15 
\end{thebibliography}
\end{document}